\documentclass[pdftex,twocolumn,epjc3]{svjour3}
\smartqed
\RequirePackage{graphicx, amsmath, amssymb, booktabs, array}
\RequirePackage{mathptmx}
\RequirePackage{flushend}
\RequirePackage[numbers,sort&compress]{natbib}
\RequirePackage[colorlinks,citecolor=blue,urlcolor=blue,linkcolor=blue]{hyperref}

\journalname{Eur. Phys. J. C}

\def\CF{\mathrm{C_F}}
\def\CFsq{\mathrm{C_F^2}}
\def\CFcub{\mathrm{C_F^3}}
\def\CFfour{\mathrm{C_F^4}}

\def\CA{\mathrm{C_A}}
\def\CAsq{\mathrm{C_A^2}}
\def\CAcub{\mathrm{C_A^3}}

\newcommand{\TF}{\mathrm{T}_f}

\def\W{\mathcal{W}}
\def\oW{\overline{\mathcal{W}}}

\def\A{\mathcal{A}} \def\Ab{\bar{\mathcal{A}}}
\def\B{\mathcal{B}} \def\Bb{\bar{\mathcal{B}}}
\def\C{\mathcal{C}}
\def\F{\mathcal{F}}

\def\cS{\mathcal{S}}
\def\cR{\mathcal{R}}
\def\G{\mathcal{G}}

\def\K{\mathcal{K}}
\def\bA{\mathbf{A}}

\newcommand{\fA}{\mathfrak{A}}

\def\S{\Sigma}
\renewcommand{\O}{\Omega}
\newcommand{\Ob}{\bar{\Omega}}
\def\inn{\mathrm{in}}
\def\out{\mathrm{out}}
\def\X{{\scriptscriptstyle X}}
\def\R{{\scriptscriptstyle\mathrm{R}}}
\def\V{{\scriptscriptstyle\mathrm{V}}}
\def\L{{\scriptscriptstyle\mathrm{L}}}
\def\P{{\scriptscriptstyle\mathrm{P}}}

\renewcommand{\d}{\mathrm{d}}
\newcommand{\as}{\alpha_s}
\newcommand{\asb}{\bar{\alpha}_s}

\def\kt{$k_t\, $ }
\def\aktt{\mathrm{ak_t}}
\def\ktt{\mathrm{k_t}}

\def\ng{\text{ng}}
\def\cl{\text{cl}}

\def\Uh{\hat{\mathcal{U}}}
\def\uh{\hat{u}}
\def\cR{c_{\scriptscriptstyle{R}}}
\begin{document}

\title{Dijet mass up to four loops with(out) ${\boldsymbol k}_{\boldsymbol t}$ clustering}

\author{K. Khelifa-Kerfa\thanksref{e1,addr1}}

\thankstext{e1}{e-mail: kamel.khelifakerfa@univ-relizane.dz}

\institute{Department of Physics, Faculty of Science and Technology \\
	Relizane University, Relizane 48000, Algeria \label{addr1}
}

\date{ }

\maketitle

\begin{abstract}
We compute the invariant mass of dijets produced in $e^+ e^-$ annihilation processes up to four loops in perturbation theory for both anti-$k_t$ and $k_t$ jet algorithms. The calculations, performed within the eikonal approximation and employing strong-energy ordering, capture the full analytic structure of the leading Abelian and non-Abelian non-global logarithms, including full colour and jet-radius dependence. We evaluate the significance of these logarithms and the convergence of the four loop perturbative expansion by comparing with all-orders numerical results.
\end{abstract}

\section{Introduction}
\label{sec:intro}

In high-energy particle collisions, such as $e^+ e^-$ annihilation into dijets, jets are central to understanding the underlying dynamics of QCD. Jets, which represent collimated sprays of final-state hadrons produced from the fragmentation of quarks and gluons, are indispensable tools for probing QCD in both perturbative and non-perturbative regimes. They are typically reconstructed using jet algorithms, such as the anti-$k_t$ \cite{Cacciari:2008gp} and $k_t$ \cite{Catani:1993hr, Ellis:1993tq} algorithms, both of which have substantial implications for theoretical calculations and experimental measurements.

In the perturbative QCD context, calculations of jet shapes and substructure present considerable complexity, even in relatively simple leptonic processes. This is due to the presence of multiple energy scales that induce large logarithms, which spoil the perturbative expansion and necessitate resummation. These logarithms fall into two primary categories: global (abelian) logarithms (GLs), arising in QCD observables defined over the entire phase space, and non-global (non-abelian) logarithms (NGLs) \cite{Dasgupta:2001sh, Dasgupta:2002bw}, which emerge in observables defined over a restricted phase space region. While global observables have been analytically computed up to next-to-next-to-leading order (NNLO) (see, e.g., \cite{GehrmannDeRidder:2007hr, DelDuca:2016ily}) and (analytically and/or numerically) resummed to next-to-next-to-leading logarithmic (NNLL) accuracy for general observables, and even to N$^3$LL for certain cases (see, for instance, \cite{deFlorian:2004mp, Becher:2008cf,Abbate:2010xh, Chien:2010kc, Monni:2011gb, Mateu:2013gya, Hoang:2014wka, Banfi:2014sua, Banfi:2016zlc, Frye:2016okc, Frye:2016aiz, Tulipant:2017ybb, Moult:2018jzp, Bell:2018gce, Banfi:2018mcq,Procura:2018zpn, Arpino:2019ozn, Becher:2019avh, Bauer:2020npd, Kardos:2020gty, Anderle:2020mxj, Dasgupta:2022fim, Duhr:2022yyp, vanBeekveld:2024wws}), non-global observables have, until recently, been resummed only up to NLL, numerically and within the large-$N_c$ limit \cite{Dasgupta:2001sh, Dasgupta:2002bw, Banfi:2002hw}. Recent advancements have improved this to include finite-$N_c$ effects at NLL \cite{Hatta:2013iba, Hagiwara:2015bia, Hatta:2020wre, DeAngelis:2020rvq, Hamilton:2020rcu, vanBeekveld:2022zhl} as well as NNLL resummation in the large-$N_c$ limit \cite{Banfi:2021owj, Banfi:2021xzn, Becher:2023vrh, Becher:2021urs, FerrarioRavasio:2023kyg}. Most of these major advancements have been incorporated into the \texttt{PanScales} parton shower (see Ref. \cite{vanBeekveld:2023ivn} and references therein).

Fixed-order (FO) calculations of jet shapes and substructure are essential complements to all-orders resummations. FO methods typically offer analytical insights, enhancing understanding of non-global observable distributions at high energy scales. They also incorporate higher-order corrections that improve precision by accounting for effects like multiple emissions, interference terms, and virtual corrections, which may not be fully captured by resummation. Combining FO and all-orders results enables more reliable predictions across a wider spectrum of the observable, providing refined theoretical uncertainties. Since their initial computation at two loops for the hemisphere mass \cite{Dasgupta:2001sh}, NGLs have been studied for various observables (e.g., \cite{Dasgupta:2002bw, Dasgupta:2006ru, Banfi:2008qs, Banfi:2010pa, Khelifa-Kerfa:2011quw, Kerfa:2012yae, Dasgupta:2012hg, Bouaziz:2022tik}). For certain cases, higher-order terms have been computed up to twelve-loops at large-$N_c$ \cite{Schwartz:2014wha, Caron-Huot:private, Benslama:2020wib}, with recent progress, at finite-$N_c$, extending (fully) to four loops and (partially) to five-loops for a range of observables in processes such as $e^+ e^-$ annihilation \cite{Khelifa-Kerfa:2015mma, Benslama:2023gys} and Higgs/vector boson plus jet production \cite{Khelifa-Kerfa:2024udm}.

The aforementioned calculations were primarily carried out for the anti-$k_t$ jet algorithm. When alternative clustering algorithms, such as the $k_t$ algorithm, are employed, these calculations become increasingly intricate and complex. It was first observed in \cite{Appleby:2002ke, Appleby:2003sj} for the interjet energy flow that the magnitude of NGLs diminishes due to the $k_t$ algorithm’s non-linear clustering condition, which influences the secondary correlated emissions responsible for NGLs. Furthermore, \cite{Banfi:2005gj} later demonstrated that $k_t$ clustering leads to a new hierarchy of large logarithmic terms for abelian primary emissions, termed clustering logarithms (CLs) (or equivalently abelian NGLs). Both of these effects have since been validated through calculations for various non-global observables (see, e.g., \cite{Delenda:2006nf, Kelley:2012kj, Banfi:2010pa, Khelifa-Kerfa:2011quw, Kelley:2012zs, Kerfa:2012yae, Ziani:2021dxr, Bouaziz:2022tik}). Fixed-order (FO) calculations of CLs beyond the leading two loops level were first performed in \cite{Delenda:2006nf} (up to four loops) for interjet energy flow and subsequently for the single-jet-mass \cite{Delenda:2012mm} and azimuthal decorrelation (up to three loops) \cite{Benslama:2023gys}, all in $e^+ e^-$ annihilation processes and at the leading-logarithmic level, with the exception of Ref. \cite{Kerfa:2012yae}, which calculated next-to-leading logarithms at two loops. No comparable calculations currently exist for hadron-hadron collisions, an area to be addressed in a forthcoming paper \cite{VHjet_kt_4loop}.
On another front, all-orders resummation of leading CLs exists only in numerical form, as implemented in the Monte Carlo (MC) code of \cite{Dasgupta:2001sh} (limited to the $k_t$ algorithm), and more recently within the framework of Soft and Collinear Effective Theory (SCET) for both $k_t$ and Cambridge/Aachen \cite{Dokshitzer:1997in, Wobisch:1998wt} clustering algorithms \cite{Becher:2023znt}.

In this paper, we examine the leptonic process of $e^+ e^-$ annihilation into two final-state jets in the threshold limit, where the jets are produced back-to-back. We compute the invariant mass (squared) of the dijet system within the framework of eikonal theory, imposing a strong energy-ordering condition on the emitted soft gluons. This approach allows us to utilise the squared eikonal amplitudes derived in \cite{Delenda:2015tbo}. The calculations are performed for both the anti-$k_t$ and $k_t$ jet algorithms. Our observable is closely related to that studied in \cite{Kelley:2011tj, Khelifa-Kerfa:2011quw, Kerfa:2012yae}, with the key difference that we impose no energy cutoff on the soft activity outside the two jets, thereby eliminating any logarithms associated with this cutoff. Furthermore, this paper extends the calculations of these prior studies, which were limited to two loops, up to four loops.

For anti-$k_t$ clustering, our calculations closely follow those conducted for the hemisphere mass observable \cite{Khelifa-Kerfa:2015mma}. In the case of the dijet mass, however, there is a dependence on the jet-radius $R$, affecting both the argument of the large non-global logarithms and the coefficients that multiply them at each order in the perturbative expansion. Calculations are carried out analytically wherever feasible; otherwise, numerical approximations are employed. Notably, in the anti-$k_t$ algorithm, no clustering logarithms arise, as this clustering algorithm tends to produce more collimated jets (similar to cone algorithms), rendering it less sensitive to soft radiation located far from the jet axis.

The $k_t$ jet algorithm, by contrast, is more inclusive, clustering soft radiation earlier in the recombination sequence. This leads to a reorganisation of the final-state partons, altering the original phase space and resulting in the previously mentioned effects: a reduction in the size of NGLs (through the clustering of small-angle secondary radiation) and the emergence of CLs (due to the clustering of soft radiation into the jet, thereby increasing its effective size \cite{Becher:2023znt}). Unlike CLs, which were computed up to four loops some time ago, as discussed previously, NGLs in $k_t$ clustering have been computed only up to two loops. However, very recently, the three loop coefficient was successfully determined for the dijet azimuthal decorrelation observable in $e^+ e^-$ annihilation \cite{Benslama:2023gys}. In this work, we employ the recently derived master formula (Eq. (21)) from Ref. \cite{Khelifa-Kerfa:2024roc} to compute NGLs up to four loops for the $k_t$ jet algorithm. This formula systematises and generalises the calculation for any non-global observable in both lepton and hadron processes. Due to the complex phase space factors in the relevant integrals, these calculations are carried out numerically; nevertheless, they include full dependence on both jet-radius and colour.

We compare the two loop distributions for both anti-$k_t$ and $k_t$ with the fixed-order next-to-leading order (NLO) Monte Carlo (MC) program \texttt{event2} \cite{Catani:1996jh}, interfaced with the \texttt{Fastjet} package \cite{Cacciari:2011ma}, and find consistency within the precision of our calculations. Furthermore, a pattern of exponentiation for NGLs and CLs emerges clearly up to four loops. It is instructive to separately compare the exponentiated NGLs and CLs to the output of the MC code from \cite{Dasgupta:2001sh} to examine the convergence of the perturbative series. Our results indicate that the inclusion of higher-order terms expands the overlap range between the analytical exponentiation and the numerical results. Notably, while a comparison with all-orders resummation at finite colour would be preferable, such resummation is currently only available, in $e^+ e^-$ processes, for the hemisphere mass observable \cite{Hatta:2013iba, Hagiwara:2015bia}.
We also derive the all-orders resummed formula, incorporating the primary Sudakov form factor, the exponential of NGLs, and the exponential of CLs (in the $k_t$ clustering case), to evaluate the significance of the latter two effects in phenomenological studies. Our findings confirm previous conclusions, specifically: (a) NGLs can reduce the Sudakov form factor by up to
$25\%$ near the peak of the distribution, and (b) the CLs form factor counteracts the NGLs effect, moderating the Sudakov factor reduction to approximately $5\%$, and even down to only $2\%$ for larger jet-radii.

This paper is organised as follows. In Sec. \ref{sec:Defn}, we present the definitions of the various components, including the jet shape observable and the jet algorithms. Sec. \ref{sec:FO} is dedicated to detailed calculations of the dijet mass at one-, two, three and four loops in the anti-$k_t$ algorithm, with comparisons to \texttt{event2} shown in the same section. The corresponding calculations for the $k_t$ algorithm are presented in Sec. \ref{sec:FOKT}. Sec. \ref{sec:Resummation} presents the all-orders analytical calculations, including a quantification of the contributions of NGLs and CLs to the full resummation. In the same section, we compare the exponentiated NGLs and CLs up to four loops with the numerical distribution from the MC code of \cite{Dasgupta:2001sh}, assessing the effect of higher-order terms. Finally, we summarise our findings and outline future directions in Sec. \ref{sec:Conclusion}.

\section{Definitions}
\label{sec:Defn}

Consider the simple QCD process of \( e^+ e^- \) annihilation into dijets that are produced, in the threshold limit, back-to-back in the lab frame. At the partonic level, we may write:
\begin{align}\label{eq:Def:AnnihilationProcess}
 e^+ + e^- \to q(p_a) + \bar{q}(p_b) + g_1(k_1) + \cdots + g_n(k_n),
\end{align}
where the quantities in parentheses denote the four-momenta of the respective partons (the quark \( q \), the anti-quark \( \bar{q} \), and gluons \( g_{i}, i = 1, \dots, n \)). These can be expressed in polar parametrisation (spherical coordinates) as follows:
\begin{align}\label{eq:Def:Momenta}
 p_a &= \frac{Q}{2} \left(1,0,0,1\right), \notag\\
 p_b &= \frac{Q}{2} \left(1,0,0,-1\right), \notag\\
 k_i &= \omega_i \left(1,s_i \cos\phi_i,s_i \sin\phi_i,c_i\right),
\end{align}
where \( c_i \equiv \cos\theta_i \) and \( s_i \equiv \sin\theta_i \), with \( \theta_i \) and \( \phi_i \) being the polar and azimuthal angles of the \( i^{\text{th}} \) gluon. The energy scales \( Q \) and \( \omega_i \) represent the hard scale of the process and the energy of the \( i^{\text{th}} \) soft gluon \( g_i \), respectively. All partons are assumed to be massless. At single-logarithmic (SL) accuracy—i.e., retaining logarithmic terms up to \( \alpha_s^n L^{2n-2} \) in the perturbative expansion of the non-global jet-mass observable—recoil effects can safely be neglected. Additionally, at SL accuracy, we work within the soft eikonal approximation and impose strong energy-ordering of the soft gluons: \( Q \gg \omega_1 \gg \omega_2 \gg \cdots \gg \omega_n \). These latter two approximations greatly simplify calculations at higher orders in the perturbative series.

The normalised invariant mass (squared) of the dijets, $\varrho$, is defined as:
\begin{align}\label{eq:Def:Rho}
 \varrho &= \varrho_{\R} + \varrho_{\scriptscriptstyle L},
 \end{align}
where
\begin{align}\label{eq:Def:Rho-R}
\varrho_\R  &= \frac{4 m_R^2}{Q^2} = \frac{4}{Q^2} \left(p_a + \sum_{i\in j_\R} k_i \right)^2 = \sum_{i \in j_\R} \varrho_{\R,i} + \mathcal{O} \left( \frac{\omega^2}{Q^2}\right)
 \notag\\
\varrho_{\R,i} &= 8\frac{p_a \cdot k_i}{Q^2} = \frac{4 \omega_i}{Q} \left(1-c_i\right) = 2 x_i \left(1 - c_i\right),
\end{align}
where \( x_i = 2\omega_i/Q \) is the energy fraction of the \( i^{\text{th}} \) gluon, and the sum runs over all soft gluons that are clustered by the jet algorithm into the right (quark-initiated) jet \( j_\R \). An identical expression holds for the normalised invariant mass of the left (anti-quark-initiated) jet \( j_{\scriptscriptstyle L} \). In the above definition of \( \varrho_{\R} \), terms proportional to \( \omega^2 / Q^2 \) are neglected in the soft limit.

The anti-$k_t$ and $k_t$ jet algorithms can be defined following the framework of Ref. \cite{Cacciari:2011ma}. In this framework, both algorithms are members of a more general class of sequential recombination algorithms known as the “generalised $k_t$ algorithm,” which depends on a continuous parameter \( p \). Specifically, this algorithm proceeds as follows:
\begin{enumerate}
	\item For each pair in the initial list of final-state particles, define the two distance measures:
	\begin{align}\label{eq:Def:DistanceMeasures}
		d_{ij} &= \min\left(E_i^{2p}, E_j^{2p}\right) \frac{1 - \cos\theta_{ij}}{1 - \cos R},
		\notag\\
		d_{iB} &= E_i^{2p},
	\end{align}
	where \( E_i \) is the energy of the \( i^{\text{th}} \) particle, \( R \) is the jet-radius parameter, and \( \cos\theta_{ij} = c_i c_j + s_i s_j \cos\phi_{ij} \) with \( \phi_{ij} = \phi_i - \phi_j \). The anti-$k_t$ and $k_t$ jet algorithms correspond to values of the parameter \( p = -1 \) and \( p = 1 \), respectively. Here, the subscript \( B \) in $d_{iB}$ refers to the beam direction.

	\item Let \( d_{\min} \) be the minimum of all \( d_{ij} \) and \( d_{iB} \). If \( d_{\min} \) corresponds to a \( d_{ij} \), then particles \( i \) and \( j \) are merged into a single pseudo-jet, with their momenta summed up (in the \( E \)-scheme recombination). If \( d_{\min} \) corresponds to a \( d_{iB} \), then particle \( i \) is declared a final (inclusive) jet and removed from the list of final-state particles.

	\item Steps 1 and 2 are repeated until no particles remain in the list.
\end{enumerate}
Notice that for the anti-$k_t$ jet algorithm (\( p = -1 \)), clustering proceeds with the harder particles first, as they minimise the energy-reciprocal measure \( E^{-2} \). If no other particles are within a distance \( R \) from a given hard particle, then that particle forms an inclusive jet. Consequently, hard, well-separated jets appear as circular regions on the \( \theta\text{-}\phi \) plane with radius \( R \). This approach offers a simple, infrared- and collinear-safe alternative to cone-like algorithms for \( e^+ e^- \) collisions \cite{Cacciari:2005hq}.

In contrast, the $k_t$ jet algorithm (\( p = 1 \)) clusters the softest particles first, as they have the minimum energies \( E^2 \). Two soft particles are merged if their distances satisfy:
\begin{align}\label{eq:Def:ClusteringCondition}
	1 - \cos\theta_{ij} < 1 - \cos R.
\end{align}
Thus, a soft gluon initially located outside a jet (perhaps during the early stages of the $k_t$ jet algorithm) can later be clustered with a harder gluon already inside the jet. This results in the soft gluon being “dragged” into the jet, where it contributes to the jet-mass. Such behaviour is absent in the anti-$k_t$ algorithm and demonstrates that clustering among soft gluons can significantly impact the jet-mass distribution. It also highlights that clustering among soft gluons is inherently complex, implying that their contribution to the jet-mass observable will be equally intricate. This observation will be further substantiated later in the text.

In this paper, we focus on calculating the jet-mass integrated cross-section (or, equivalently, the jet-mass cumulant), \( \Sigma(\rho) \), defined as:
\begin{align}\label{eq:Def:IntegratedJetMass}
	\Sigma(\rho) = \int \frac{1}{\sigma_0} \frac{\d \sigma}{\d \varrho} \, \Theta\left[\rho - \varrho(k_1,\dots,k_n)\right] \, \Xi(k_1, \dots, k_n) \, \d \varrho,
\end{align}
where \( \sigma_0 \) denotes the Born cross-section, and \( \d\sigma / \d\varrho \) represents the differential cross-section in the normalised jet-mass \( \varrho(k_1, \dots, k_n) \), which is restricted to be less than a threshold \( \rho \). The clustering function \( \Xi(k_1, \dots, k_n) \) arises from the application of the chosen jet algorithm and limits the final-state phase space to only include gluonic configurations that contribute to the jet-mass. The fixed-order perturbative expansion of this integrated cross-section can be expressed as:
\begin{align}
	\Sigma(\rho) = 1 + \Sigma_1(\rho) + \Sigma_2(\rho) + \cdots,
\end{align}
where the \( m^{\text{th}} \) term in the expansion is given by:
\begin{multline}\label{eq:Def:Sigma-m}
	\Sigma_m(\rho) = \sum_{X} \int_{x_1 > x_2 > \cdots > x_m} \prod_{i=1}^m \d \Phi_i \times \\ \times  \Uh_m\, \mathcal{W}^X_{1 \dots m} \, \Xi_m(k_1, \dots, k_m),
\end{multline}
where the phase-space factor for the emission of the \( i^{\text{th}} \) soft gluon is given by:
\begin{align}
	\d \Phi_i = \bar{\alpha}_s \, \frac{\d x_i}{x_i} \, \d c_i \frac{\d \phi_i}{2 \pi},
\end{align}
with \( \bar{\alpha}_s = \alpha_s / \pi \). The eikonal amplitude squared for the emission of \( m \) soft, energy-ordered gluons in the configuration \( X \), normalised to the Born cross-section, \( \mathcal{W}^X_{1 \dots m} \), has been defined and extensively discussed in Ref. \cite{Delenda:2015tbo} for \( e^+ e^- \) processes. Each gluon may be real (R) or virtual (V), and \( X \) denotes a possible configuration of \( m \) such gluons. For example, \( X \) could be \( \{RR \dots V\} \), where \( g_1 \) and \( g_2 \) are real, \( \dots \), and \( g_m \) is virtual. The summation in Eq.~\eqref{eq:Def:Sigma-m} runs over all possible configurations \( X \) at each order in the perturbative series. Explicit formulae for the eikonal amplitudes squared up to four loops are provided in the aforementioned reference \cite{Delenda:2015tbo}.

The measurement operator at the \( m^{\text{th}} \)-order, \( \Uh_m \), acts on a given eikonal amplitude squared to ensure that only gluon configurations for which the jet-mass observable is less than \( \rho \) contribute non-trivially to the integrated cross-section \( \S(v) \). All other configurations are set to zero. This operator was first introduced by Schwartz and Zhu in \cite{Schwartz:2014wha} and later utilised in \cite{Khelifa-Kerfa:2015mma} for calculating the hemisphere mass observable up to four loops in the anti-$k_t$ algorithm. Interested readers are referred to these references for comprehensive details. Notably, in both the anti-$k_t$ and $k_t$ algorithms, it factorises as follows \cite{Schwartz:2014wha, Khelifa-Kerfa:2015mma, Khelifa-Kerfa:2024roc}:
\begin{align}\label{eq:Def:Uh}
	\Uh_m = \prod_{i=1}^m \uh_i,
\end{align}
where the measurement operator for the \( i^{\text{th}} \) emission reads:
\begin{align}\label{eq:Def:uh}
	\uh_i = \Theta_i^\V + \Theta_i^\R \left[\Theta_i^\out + \Theta_i^\inn \Theta\left(\rho - \varrho_i\right) \right] = 1 - \Theta_i^\R \Theta_i^\inn \Theta_i^\rho,
\end{align}
with the step-function \( \Theta_i^{\V(\R)} \) equals to one if gluon \( g_i \) is virtual (real) and zero otherwise. Additionally, \( \Theta_i^{\out(\inn)} \) is equal to one if gluon \( g_i \) is outside (inside) either jet \( j_\R \) or \( j_\L \) (after applying the jet algorithm) and zero otherwise. The jet-mass step-function is defined as \( \Theta_i^\rho \equiv \Theta\left( \varrho_i - \rho\right) \). Note that \( \Theta_i^\R + \Theta_i^\V = 1 \) and \( \Theta_i^\inn + \Theta_i^\out = 1 \).

We begin by presenting fixed-order calculations in the anti-$k_t$ jet algorithm in the next section, Sec.~\ref{sec:FO}, and then proceed to the $k_t$ jet algorithm in Sec.~\ref{sec:FOKT}.

\section{Fixed-order calculations in anti-\kt}
\label{sec:FO}

\subsection{One and two loops}
\label{sec:1-2loop}

\begin{figure}[h!]
\centering
\includegraphics[scale=0.57]{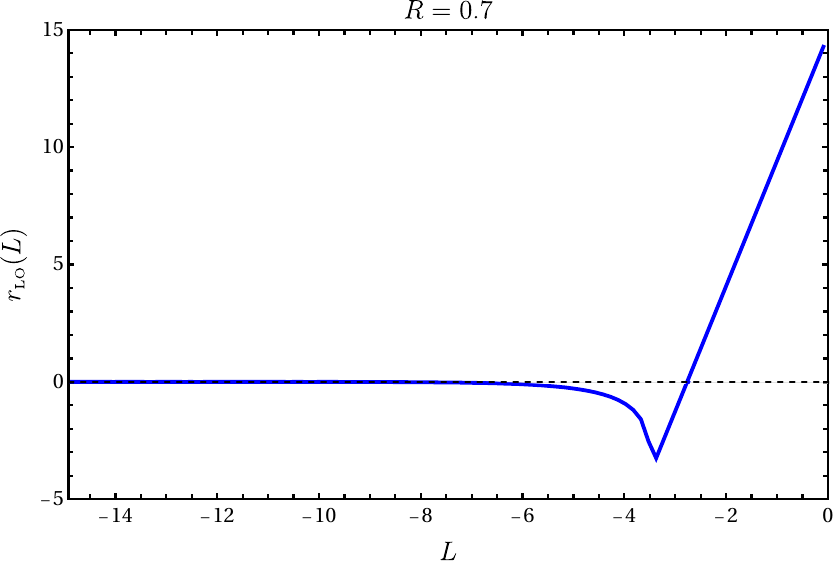}
\caption{Comparisons between the analytical and \texttt{event2} differential distributions for the dijet mass observable at one-loop.} \label{fig:1loop}
\end{figure}

At one-loop, the quark–anti-quark hard antenna emits a single soft gluon \( k_1 \), which can be either real or virtual. The corresponding eikonal amplitudes squared are thus \cite{Delenda:2015tbo}:
\begin{align}\label{eq:1loop:EikAmp}
	\W_1^\R = \CF\,w_{ab}^1, \qquad \W_1^\V = - \W_1^\R,
\end{align}
where the Casimir colour factor \( \CF = (N_c^2-1)/2N_c \), with \( N_c \) being the number of colours in the fundamental representation, and the one-loop antenna function reads:
\begin{align}\label{eq:1loop:AntennaFunc}
	w_{ab}^k = \omega_k^2\, \frac{\left(p_a \cdot p_b\right)}{ \left(p_a \cdot k\right) \left(k \cdot p_b\right)  }
	= \omega_k^2\, \frac{(ab)}{(ak)(kb)}.
\end{align}
Summing over these two gluon configurations, we find for the integrand in \eqref{eq:Def:Sigma-m}
\begin{align}\label{eq:1loop:uWX}
	\sum_\X \uh_1 \W_1^\X = \uh_1 \W_1^\R + \uh_1 \W_1^\V = -\Theta_1^\rho \Theta_1^\inn\, \W_1^\R,
\end{align}
where we have applied the one-loop measurement operator \( \uh_1 \) from \eqref{eq:Def:uh} on the real and virtual eikonal amplitudes squared to obtain the last equality. We note that at one-loop order, there are no differences between the various jet algorithms, and they all yield the same results. From the above equation \eqref{eq:1loop:uWX}, it is straightforward to see that the anti-$k_t$ clustering function is, at one-loop, simply equal to:
\begin{align}\label{eq:1lop:ClusFunc}
	\Xi_1^{\aktt}(k_1) = \Theta_1^\inn,
\end{align}
where the superscript ``$\aktt$" is a shorthand for the anti-$k_t$ jet algorithm. Substituting \eqref{eq:1loop:uWX} into the formula for the cumulant \eqref{eq:Def:Sigma-m} at one-loop gives:
\begin{align}\label{eq:1loop:Sigma}
	\S_1^{\aktt}(\rho) &= -\int \d\Phi_1\, \Theta_1^\rho \Theta_1^\inn\, \W_1^\R,
	\notag\\
	&=  -\CF\,\asb\, \int_0^1 \frac{\d x_1}{x_1} \Big(\int_{-1}^{-\cR} \Theta\left[2x_1(1+c_1) - \rho\right] + \notag\\
	&+ \int_{\cR}^1 \Theta\left[2x_1(1-c_1) - \rho\right] \Big) \d c_1 \int_0^{2\pi} \frac{\d \phi_1}{2\pi}\, w_{ab}^1,
\end{align}
where \( \cR \equiv \cos R \), and \( \Theta_1^\inn \) has been replaced by the expression \( \Theta\left(1-c_1, c_1 - \cR\right) + \Theta\left(1+c_1, -\cR - c_1\right) \). The one-loop antenna function \( w_{ab}^1 \) is given by \( 2/(1 - c_1^2) \). To capture the hard-collinear contribution to the distribution, we replace \( 1/x_1 \) with \( (1 + (1 - x_1)^2)/2x_1 \). Performing the integration up to SL accuracy, we obtain:
\begin{align}\label{eq:1loop:Sigma-Final}
	\S_1^{\aktt}(\rho) = -\CF\,\asb \left[L^2 + \left(\frac{3}{2} - 2\ln\frac{1-\cR}{1+\cR} \right)\, L \right] \equiv \S_1^\P(\rho),
\end{align}
where \( L = \ln(1/\rho) \) and the superscript \( \mathrm{P} \) refers to primary emissions. That is, the leading logarithms in the dijet mass distribution are entirely captured by primary emissions off the hard \( q \bar{q} \) pair. They originate from soft and collinear singularities of the corresponding part of the eikonal amplitude squared. The all-orders resummation of these leading double logarithms, which form the Sudakov form factor, has been known for decades as a simple exponential of the one-loop result \eqref{eq:1loop:Sigma-Final}. Figure \ref{fig:1loop} shows the difference between the analytical and exact (\texttt{event2}) differential distributions at leading order for a jet-radius \( R = 0.7 \):
\begin{align}
	r_{\scriptscriptstyle\mathrm{LO}}(L) = \frac{\d\sigma^{\texttt{event2}}_{\text{LO}}}{\d \tilde{L}} -  \frac{\d\S_1^{\aktt}}{\d \tilde{L}},
\end{align}
where \( \tilde{L} = \ln \rho = -L \) and the factor \( (\as/2\pi) \) is omitted \footnote{The differential distributions in \texttt{event2} exclude the prefactor \( (\as/2\pi) \) and the Casimir colour factors, at both leading- and next-to-leading order.}. As expected, the difference tends to zero at large logarithms. The cutoff in the \texttt{event2} distribution corresponds to the maximum value of the dijet mass observable, given by \cite{Khelifa-Kerfa:2011quw, Kerfa:2012yae}:
\begin{align}
	\rho_{\max} = \frac{8}{R^2} \left(1 - \sqrt{1 - R^2/4}\right) - 1.
\end{align}
We emphasise again that Fig.~\ref{fig:1loop} would be identical for all jet algorithms since, as mentioned above, they operate equivalently at this order. It is also worth noting that the result in \eqref{eq:1loop:Sigma-Final} matches that found in \cite{Khelifa-Kerfa:2011quw, Kerfa:2012yae}, corresponding to the \( \tau_{E_0} \) part of the distribution with \( 1 - \cR = 2 R_s \).

For the emission of two soft, energy-ordered gluons, \( k_1 \) and \( k_2 \), the sum over all possible gluon configurations \eqref{eq:Def:Sigma-m} at two loop order is given by \cite{Khelifa-Kerfa:2015mma, Khelifa-Kerfa:2024roc}:
\begin{align}\label{eq:2loop:uWX-A}
	\sum_{\X} \Uh_2 \W_{12}^\X &= \uh_1 \uh_2 \left(\W_{12}^{\R\R} + \W_{12}^{\R\V} + \W_{12}^{\V\R} + \W_{12}^{\V\V}\right), \notag\\
	&= -\Theta_1^\rho \Theta_2^\rho \Theta_2^\inn \left[\W_{12}^{\V\R} + \Theta_1^\out \W_{12}^{\R\R}\right],
\end{align}
where the second equality arises from applying the measurement operators of the first and second emissions on the various eikonal amplitudes squared. Note that other gluon configurations either do not produce real-virtual mis-cancellations (and thus no large logarithms) or do produce large logarithms but are subleading. For instance, configurations where the harder gluon \( k_1 \) is inside the left-jet \( j_\L \) and radiates the softer gluon \( k_2 \) inside the right-jet \( j_\R \). Although the softer gluon \( k_2 \) contributes to the mass of \( j_\R \), this contribution is subleading because \( k_1 \), being inside \( j_\L \), is vetoed. Such scenarios are discussed in detail in Refs.~\cite{Khelifa-Kerfa:2011quw, Kerfa:2012yae}.

Using the relation \( \Theta_1^\inn + \Theta_1^\out = 1 \) and the explicit expressions of the eikonal amplitudes squared \cite{Delenda:2015tbo}, we can further simplify the above equation to:
\begin{align}\label{eq:2loop:uWX-B}
	\sum_{\X} \Uh_2 \W_{12}^\X  &= -\Theta_1^\rho \Theta_2^\rho \left[-\Theta_1^\inn \Theta_2^\inn  \W_1^\R \W_2^\R + \Theta_1^\out \Theta_2^\inn \oW_{12}^{\R\R}\right],
\end{align}
where the one-loop eikonal amplitude squared, \( \W_i^\R \), is given above, and the \textit{irreducible} part of the two loop eikonal amplitude squared, \( \oW_{12}^{\R\R} \), reads \cite{DelDuca:2016ily}:
\begin{align}\label{eq:2loop:W2Ir}
	\oW_{12}^{\R\R} = \frac{1}{2} \CF \CA\, \A_{ab}^{12}, \qquad
	\A_{ab}^{ij} = w_{ab}^i \left(w_{ai}^j + w_{ib}^j - w_{ab}^j \right).
\end{align}
The term \( \A_{ab}^{ij} \) is known as the two loop antenna function. From Eq.~\eqref{eq:2loop:uWX-B} we can see that the two loop clustering function for primary emissions is simply the product of the corresponding one-loop functions \eqref{eq:1lop:ClusFunc}, whereas, for secondary correlated emissions, it is given by:
\begin{align}\label{eq:2loop:ClusFunc}
	\Xi_{2}^{\aktt} (k_1, k_2) = \Theta_1^\out \Theta_2^\inn.
\end{align}
Substituting Eq.~\eqref{eq:2loop:uWX-B} into the expression of the cumulant \eqref{eq:Def:Sigma-m} at two loops, we obtain:
\begin{align}\label{eq:2loop:Sigma}
	\S_2(\rho) = \int_{1_\inn} \int_{2_\inn} \W_1^\R \W_2^\R - \int_{1_\out} \int_{2_\inn} \oW_{12}^{\R\R},
\end{align}
where we have introduced the shorthand notation:
\begin{align}\label{eq:Def:IntShorthandNotation}
	\int_{i_y} = \int \d\Phi_i \, \Theta_i^\rho\, \Theta_i^y, \qquad y \in \{\inn, \out\},
\end{align}
and strong-energy ordering \( x_1 > x_2 \) is implied. The first part on the right-hand side (rhs) of Eq.~\eqref{eq:2loop:Sigma} is the two loop term in the series expansion of the Sudakov form factor:
\begin{align}
	\int_{1_\inn} \int_{2_\inn} \W_1^\R \W_2^\R  = \frac{1}{2!} \left(\S_1^\P\right)^2.
\end{align}
The second part on the rhs of Eq.~\eqref{eq:2loop:Sigma} represents the first pure NGLs contribution to the dijet mass distribution. At SL accuracy, the energy integrals factorise out, resulting in a logarithm \( L^2 / 2! \) (see, for instance, \cite{Khelifa-Kerfa:2015mma} for details). The remaining integral can be cast as:
\begin{align}\label{eq:2loop:SigmaNG}
	\S_{2,\ng}^{\aktt}(\rho) = -2\,\asb^2\,\frac{L^2}{2!}\,\CF\CA\, \G_2^{\aktt}(R),
\end{align}
where the factor of 2 accounts for both jets, \( j_\R \) and \( j_\L \), and the two loop NGLs coefficient \( \G_2^{\aktt}(R) \) is given by:
\begin{align}\label{eq:2loop:G2}
\G_2^{\aktt}(R) &= \frac{1}{2} \int_{1_\out} \int_{2_\inn} \A_{ab}^{12} = \zeta_2 - \mathrm{Li}_2 \left[\left( \frac{1-\cR}{1+\cR}\right)^2\right].
\end{align}
In the above expression for $\G_2^{\aktt}(R)$, it is understood that the integrals are purely angular and therefore do not involve the term $\asb^2 \d/\d x_1 \d/\d x_2\, \Theta_1^\rho \Theta_2^\rho$, as the latter has already been integrated out to yield the $\asb^2 L^2 / 2!$ term. Although we use the same shorthand notation, it should be clear throughout the paper that, whenever the integral notation $\int_{i}$ is employed for NGLs and/or CLs coefficients, the integrals are purely angular.

The analytical formula \eqref{eq:2loop:G2} is equivalent to the result in \cite{Dasgupta:2002bw, Appleby:2002ke, Khelifa-Kerfa:2011quw, Kerfa:2012yae}, expressed in terms of the rapidity gap \( \Delta \eta \), which is related to the jet-radius \( R \) by \( \Delta \eta = -\ln(1-\cR)/(1+\cR) \). Figure \ref{fig:F2G2} shows a plot of \( \G_2(R) \) as a function of \( R \) for both anti-$k_t$ and $k_t$ algorithms; we defer discussion of these curves to Sec.~\ref{sec:2loopKT}. In the limit \( R \to 0 \), one recovers the hemisphere mass result \cite{Dasgupta:2001sh, Khelifa-Kerfa:2015mma}: \( \G_2^{\aktt} = \zeta_2 \).

In Figure \ref{fig:2loop}, we plot the difference between the analytical and \texttt{event2} differential distributions at two loop for a jet-radius \( R = 0.7 \) and for all three colour channels; \( \CFsq, \CF\CA \) and \( \CF \TF\, n_f \):
\begin{equation}\label{eq:2loop:rNLO}
	r_{\mathrm{NLO}}(L) = \frac{\d \sigma^{\texttt{event2}}}{\d \tilde{L}} - \frac{\d \S_2^{\aktt}}{\d \tilde{L}}.
\end{equation}
The details of the analytical distribution are found in Appendices A, B, and C of Ref.~\cite{Khelifa-Kerfa:2011quw} \footnote{The distribution of the dijet mass observable may be derived from the jet-thrust \( \tau_{E_0} \) of \cite{Khelifa-Kerfa:2011quw} up to two loops by setting \( \log(2E_0/Q) \) in \cite{Khelifa-Kerfa:2011quw} to zero.}. We observe that all three curves tend to a constant at large values of \( \tilde{L} \), indicating agreement up to \( \asb^2 L \) in the differential distribution. Notice that the results of \texttt{event2} were obtained by running the program for $10^{11}$ events.

\subsection{Three loops}
\label{sec:3loop}

In the case of the emission of three soft, energy-ordered gluons, \( k_1 \), \( k_2 \), and \( k_3 \), the sum over all possible gluon configurations yields the formula \cite{Khelifa-Kerfa:2011quw, Khelifa-Kerfa:2024roc}:
\begin{align}\label{eq:3loop:uWX}
	\sum_\X \Uh_3 \W_{123}^\X &= -\Theta_1^\rho \Theta_2^\rho \Theta_3^\rho \Theta_3^\inn \Big[
	\W_{123}^{\V\V\R} + \Theta_1^\out \W_{123}^{\R\V\R} + \notag\\
	& + \Theta_2^\out \W_{123}^{\V\R\R} + \Theta_1^\out \Theta_2^\out \W_{123}^{\R\R\R}
	\Big].
\end{align}
Following the procedure outlined above for the two loop calculations and substituting the expressions for the three loop eikonal amplitudes squared from Ref.~\cite{Delenda:2015tbo}, we find that the cumulant \eqref{eq:Def:Sigma-m} at this order (\( m=3 \)) is given by:
\begin{align}\label{eq:3loop:Sigma-A}
	\S_3^{\aktt}(\rho) &=
	-\int_{1_\inn} \int_{2_\inn} \int_{3_\inn} \W_1^\R \W_2^\R \W_3^\R
	+ \int_{2_\inn} \W_2^\R \, \int_{1_\out} \int_{3_\inn} \oW_{13}^{\R\R}  \notag\\
	&+ \int_{3_\inn} \W_3^\R \, \int_{1_\out} \int_{2_\inn} \oW_{12}^{\R\R}
	+\int_{1_\inn} \W_1^\R\,  \int_{2_\out} \int_{3_\inn} \oW_{23}^{\R\R} \notag\\
	& + \S_{3, \ng}^{\aktt}.
\end{align}
The first term in the equation above is simply the three loop term in the expansion of the Sudakov form factor; \( \left(\S_1^\P\right)^3/3! \). The second, third, and fourth terms represent the interference terms between the one-loop primary contribution \eqref{eq:1loop:Sigma-Final} and the two loop NGLs contribution \eqref{eq:2loop:SigmaNG}, i.e., \( \S_1^\P \times \S_{2,\ng}^{\aktt} \). The final term in Eq.~\eqref{eq:3loop:Sigma-A} is the new irreducible NGLs contribution at this order, given by \cite{Khelifa-Kerfa:2011quw}:
\begin{align}\label{eq:3loop:SigmaNG-A}
	\S_{3,\ng}^{\aktt}(\rho) &= - \int_{1_\out} \int_{2} \int_{3_\inn} \left[\oW_{123}^{\R\V\R} + \Theta_2^\out \oW_{123}^{\R\R\R} \right],
\end{align}
where the subscript \( 2 \) in the second integral denotes that gluon \( k_2 \) is neither in nor out. Note that \( \int_{i_\inn} = \int_i \Theta_i^{\inn} \) and \( \int_{i_\out} = \int_i \Theta_i^{\out} \).
Using the relation \( \Theta_2^\inn + \Theta_2^\out = 1 \) along with the explicit forms of the irreducible eikonal amplitudes squared, \( \oW_{123}^{\R\V\R} \) and \( \oW_{123}^{\R\R\R} \), the above NGLs contribution can be expressed as:
\begin{align}\label{eq:3loop:SigmaNG-B}
	\S_{3,\ng}^{\aktt}(\rho) = + 2\,\asb^3 \frac{L^3}{3!}\, \CF\CAsq\, \G_3^{\aktt}(R),
\end{align}
where the factor of 2 accounts for both jets, and the three loop NGLs coefficient \( \G_3^{\aktt} \) is:
\begin{multline}\label{eq:3loop:G3}
	\G_3^{\aktt}(R) = \frac{1}{4} \left[\int_{1_\out} \int_{2_\inn} \int_{3_\inn} \A_{ab}^{12} \Ab_{ab}^{13}
	- \int_{1_\out} \int_{2_\out} \int_{3_\inn} \B_{ab}^{123}\right],
\end{multline}
with the three loop antenna function \( \B_{ab}^{ijk} \) defined in Ref.~\cite{Khelifa-Kerfa:2015mma} (Eq.~(39b)). The result of this integration as a function of the jet-radius is shown in Fig.~\ref{fig:F3G3}. It is evident from the figure that, in the limit \( R \to 0 \), one obtains \( \G_3^{\aktt} = \zeta_3 \), which aligns with the result for the hemisphere mass distribution \cite{Khelifa-Kerfa:2011quw}.

\subsection{Four loops}
\label{sec:4loop}

For the emission of four soft, energy-ordered gluons, $k_1, k_2, k_3$, and $k_4$, the calculation of the dijet mass distribution follows an analogous procedure to the two- and three loop cases. Specifically, the sum over possible gluon configurations is given by:
\begin{align}\label{eq:4loop:uWX}
	\sum_\X \Uh_4 \W_{1234}^\X &= -\prod_{i=1}^4 \Theta_i^\rho\, \Theta_4^\inn \Big[
	\W_{1234}^{\V\V\V\R} + \Theta_1^\out \W_{1234}^{\R\V\V\R} + \notag\\
	&+ \Theta_2^\out \W_{1234}^{\V\R\V\R} + \Theta_3^\out \W_{1234}^{\V\V\R\R} + \Theta_1^\out \Theta_2^\out \W_{1234}^{\R\R\V\R} + \notag\\
	&+ \Theta_1^\out \Theta_3^\out \W_{1234}^{\R\V\R\R} + \Theta_2^\out \Theta_3^\out \W_{1234}^{\V\R\R\R} + \notag\\
	&+ \Theta_1^\out \Theta_2^\out \Theta_3^\out \W_{1234}^{\R\R\R\R}
	\Big].
\end{align}
Substituting this expression into \eqref{eq:Def:Sigma-m} at four loops ($m=4$), and simplifying, we obtain the expected exponential expansion \cite{Khelifa-Kerfa:2015mma}:
\begin{align}\label{eq:4loop:Sigma}
	\S_{4}^{\aktt}(\rho) &= \frac{1}{4!} \left(\S_1^\P\right)^4
	+ \frac{1}{2!} \left(\S_1^\P\right)^2 \times \S_{2,\ng}^{\aktt}
	+ \S_1^\P \times \S_{3,\ng}^{\aktt} + \notag\\
	&+ \frac{1}{2!} \left(\S_{2,\ng}^{\aktt}\right)^2 + \S_{4,\ng}^{\aktt}.
\end{align}
The new irreducible NGLs contribution to the dijet mass distribution at this order is represented by the last term, $\S_{4,\ng}^{\aktt}$. Its explicit form is given by \cite{Khelifa-Kerfa:2015mma}:
\begin{multline}\label{eq:4loop:SigmaNG}
	\S_{4, \ng}^{\aktt}(\rho) = - \int_{1_\out} \int_{2} \int_{3} \int_{4_\inn}
	\Big[ \oW_{1234}^{\R\V\V\R}  + \Theta_2^\out \oW_{1234}^{\R\R\V\R}
	\\+  \Theta_3^\out \oW_{1234}^{\R\V\R\R} + \Theta_2^\out \Theta_3^\out \oW_{1234}^{\R\R\R\R}  \Big].
\end{multline}
Inserting the explicit expressions for the various eikonal amplitudes squared and simplifying, we can rewrite \eqref{eq:4loop:SigmaNG} in a form analogous to Eqs. \eqref{eq:2loop:SigmaNG} and \eqref{eq:3loop:SigmaNG-B}:
\begin{multline}\label{eq:4loop:SigmaNG-B}
	\S_{4, \ng}^{\aktt}(\rho) = -2\,\asb^4\, \frac{L^4}{4!}\, \Big[ \CF\CAcub\, \G_{4,a}^{\aktt}(R) + \\ + \CF\CAsq \left(\CA - 2\CF\right) \G_{4,b}^{\aktt}(R) \Big],
\end{multline}
where, for the first time in the dijet mass distribution, we observe the appearance of a finite-$N_c$ NGLs contribution, namely the term $\G_{4,b}^{\aktt}$, which is multiplied by the colour factor $(\CA -2\CF) = 1/N_c$. This colour factor vanishes in the large-$N_c$ limit, i.e., as $N_c \to \infty$ \footnote{In practice, the large-$N_c$ limit is achieved by simply setting $\CF \to \CA/2$.}.
The NGLs coefficients are given by:
\begin{align}\label{eq:4loop:G41}
 \G_{4,a}^{\aktt}(R) &= \K_1 -3\K_2 -\K_3 + \K_4,
\end{align}
where
\begin{subequations}
\begin{align}
 \K_1 &= \frac{1}{8} \int_{1_\out} \int_{2_\inn} \int_{3_\inn} \int_{4_\inn} \A_{ab}^{12} \Ab_{ab}^{13} \Ab_{ab}^{14},
 \\
 \K_2 &= \frac{1}{8} \int_{1_\out} \int_{2_\inn} \int_{3_\out} \int_{4_\inn} \A_{ab}^{12} \Bb_{ab}^{134},
 \\
 \K_3 &= \frac{1}{8} \int_{1_\out} \int_{2_\out} \int_{3_\inn} \int_{4_\inn} \fA_{ab}^{1234},
\\
\K_4 &= \frac{1}{8} \int_{1_\out} \int_{2_\out} \int_{3_\out} \int_{4_\inn} \C_{ab}^{1234}.
\end{align}
\end{subequations}
and
\begin{align}\label{eq:4loop:G42}
 \G_{4,b}^{\aktt}(R) &= \frac{1}{8} \int_{1_\out} \int_{2_\out} \int_{3_\inn} \int_{4_\inn}\, \bA_{ab}^{1234}.
\end{align}
The various antenna functions are defined in \cite{Delenda:2015tbo}.
Performing the integrations, the final results are plotted in Fig. \ref{fig:F4G4}. As in previous cases, in the limit $R \to 0$ we recover the hemisphere mass result \cite{Khelifa-Kerfa:2015mma}: $\G_{4,a}^{\aktt} = 29 \zeta_4/8$ and $\G_{4,b}^{\aktt} = - \zeta_4/2$.

\section{Fixed-order calculations in $k_t$}
\label{sec:FOKT}

This section is devoted to calculations of the dijet mass distribution when final-state jets are defined using the $k_t$ clustering algorithm. As previously stated, the influence of various jet algorithms emerges at two loops. Thus, at one-loop, this distribution is identical to that of the anti-$k_t$ algorithm, as given in Eq. \eqref{eq:1loop:Sigma-Final}. As is well established, $k_t$ clustering introduces a new tower of clustering logarithms (CLs) for primary emissions while also reducing the magnitude of non-global logarithms (NGLs) for secondary correlated emissions. Therefore, at each loop-order, up to four loops, we will compute both types of logarithms.

The work presented below builds on the findings of the recently published paper  \cite{Khelifa-Kerfa:2024roc}, in which the fixed-order analytical structure of the $k_t$ clustering was determined to all-loop orders in the perturbative expansion.

\subsection{two loops}
\label{sec:2loopKT}

For the emission of two soft, energy-ordered gluons, $k_1$ and $k_2$, the sum over all possible gluon configurations of the corresponding eikonal amplitudes squared in Eq. \eqref{eq:Def:Sigma-m}, with $k_t$ clustering turned on, is given by \cite{Khelifa-Kerfa:2024roc} (see Eq. (16)):
\begin{align}\label{eq:2loopKT:uWX}
	\sum_\X \Uh_2 \W_{12}^\X = -\Theta_1^\rho \Theta_2^\rho \Theta_2^\inn \left[\W_{12}^{\V\R} + \Theta_1^\out \Ob_{12} \W_{12}^{\R\R}\right],
\end{align}
where $\O_{ij} = \Theta\left(d_{jB} - d_{ij} \right)$ and $\Ob_{ij} = 1-\O_{ij}$, with $d_{jB}$ and $d_{ij}$ defined as distance measures in Sec. \ref{sec:Defn}. Notice that if $\O_{12} = 0$ (and hence $\Ob_{12} = 1$), the above equation reduces to its anti-$k_t$ counterpart, Eq. \eqref{eq:2loop:uWX-A}. Using the complementarity relations $\Theta_2^\inn+\Theta_2^\out = 1$ and $\O_{12}+\Ob_{12} = 1$, substituting for the eikonal amplitudes squared, and simplifying, Eq. \eqref{eq:2loopKT:uWX} reduces to:
\begin{multline}\label{eq:2loopKT:uWX-B}
	\sum_\X \Uh_2 \W_{12}^\X = -\Theta_1^\rho \Theta_2^\rho \Big[-\Theta_1^\inn \Theta_2^\inn\, \W_1^\R \W_2^\R - \\ - \Theta_1^\out \Theta_2^\inn \O_{12}\, \W_1^\R \W_2^\R +  \Theta_1^\out \Theta_2^\inn \Ob_{12}\, \oW_{12}^{\R\R}
	\Big].
\end{multline}
The first term on the rhs corresponds to the case of the anti-$k_t$ (no clustering) algorithm (identical to the first term in Eq. \eqref{eq:2loop:uWX-B}), and is fully accounted for by the exponentiation of the one-loop result (the Sudakov form factor). Indeed, terms originating from the expansion of the Sudakov form factor appear at every loop order even for the $k_t$ clustering algorithm.

\begin{figure}[h!]
	\centering
	\includegraphics[scale=0.57]{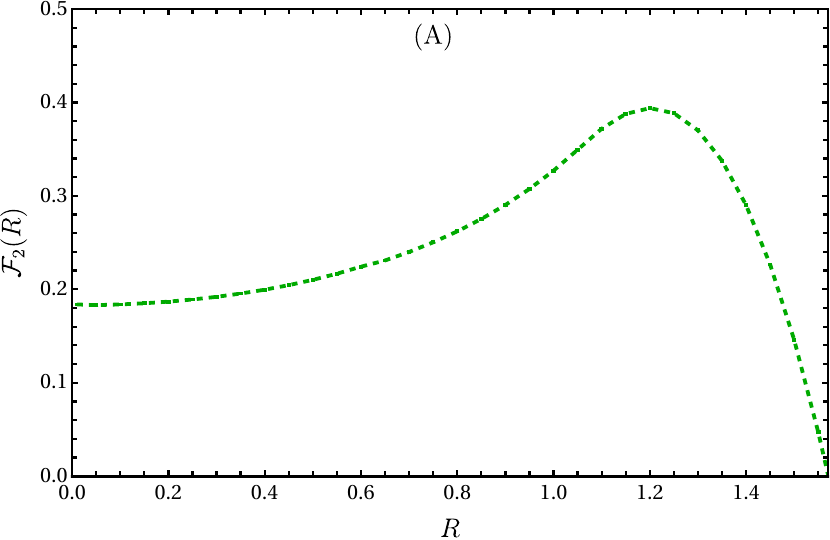} \vskip 0.5em
	\includegraphics[scale=0.57]{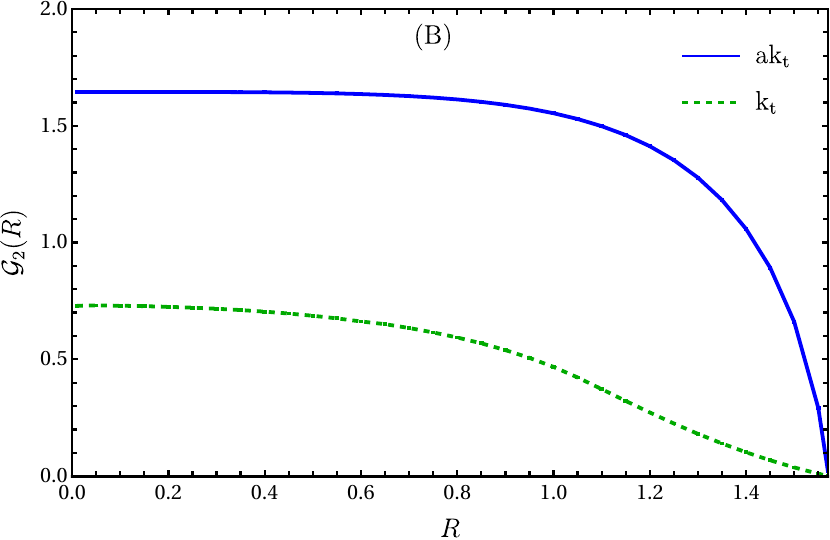} \vskip 0.5em
	\includegraphics[scale=0.54]{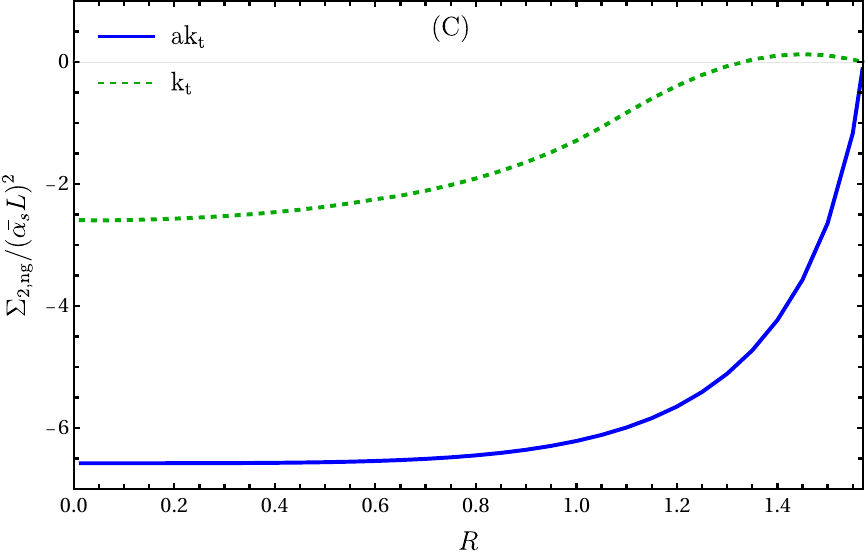}
	\caption{The two loop coefficients for CLs (A) and NGLs in both ant-$k_t$ (ak$_t$) and $k_t$ algorithms (B). A comparison between the NGLs and CLs+NGLs contributions to the dijet mass cumulant at two loops in anti-$k_t$ and $k_t$ algorithms, respectively (C).} \label{fig:F2G2}
\end{figure}

The second term in Eq. \eqref{eq:2loopKT:uWX-B} represents the new CLs contribution at this order. The corresponding clustering function can be readily identified as:
\begin{align}\label{eq:2loopKT:ClusFunc}
	\Xi_{2, \cl}^{\ktt}(k_1, k_2) = \Theta_1^\out \Theta_2^\inn\, \O_{12}.
\end{align}
This result is identical to those found in, for instance, \cite{Delenda:2006nf, Delenda:2012mm, Banfi:2010pa, Khelifa-Kerfa:2011quw}. Note that if no clustering occurs, that is, if $\O_{12} = 0$, CLs are absent. Substituting back into the cumulant expression \eqref{eq:Def:Sigma-m} for $m=2$, we find the CLs contribution:
\begin{align}\label{eq:2loopKT:Sigma-CL}
	\S_{2, \cl}^{\ktt}(\rho) = \int_{1_\out} \int_{2_\inn} \W_1^\R \W_2^\R\, \O_{12} = 2\,\asb^2\, \frac{L^2}{2!} \,\CFsq\, \F_2(R),
\end{align}
where the factor of $2$ accounts for the two jets, $L = \ln(1/\rho)$, and the two loop CLs coefficient is given by:
\begin{align}\label{eq:2loopKT:F2}
	\F_2(R) &= \int_{-\cR}^{\cR} \d c_1 \int_{\cR}^{1} \d c_2 \int_{0}^{2\pi} \frac{\d\phi_1}{2} \int_0^{2\pi} \frac{\d\phi_2}{2\pi} \, w_{ab}^1 w_{ab}^2\, \O_{12},
\end{align}
where the $k_t$ clustering condition factor, $\O_{12}$, is given explicitly by: $\O_{12} = \Theta\left(c_1 c_2 + s_1 s_2 \cos\phi_{12} - c_2\right) $. The result of the numerical integration is shown in Fig. \ref{fig:F2G2} (A). Notice that as $R \to 0$, we recover previously reported results \cite{Delenda:2006nf, Banfi:2010pa, Delenda:2012mm, Khelifa-Kerfa:2011quw}, with $\F_2(R \sim 0) = 0.183$. This is the {\it edge} or {\it boundary} effect observed also in (non-Abelian) NGLs, arising because the corresponding eikonal amplitude is most singular when the two gluons are close to each other. Thus, CLs and NGLs originate chiefly from the boundary of the measured jet \cite{Khelifa-Kerfa:2011quw, Kerfa:2012yae}. Additionally, from Fig. \ref{fig:F2G2}, we observe that the CLs coefficient increases steadily to approximately $R = 1.2$, then falls off towards zero. This may be attributed to the fact that the inter-jet region available for harder emissions decreases as jet radii increase.

The third term in Eq. \eqref{eq:2loopKT:uWX-B} represents the NGLs contribution. The corresponding clustering function reads:
\begin{align}
	\Xi_{2, \ng}^{\ktt}(k_1, k_2) = \Theta_1^\out \Theta_2^\inn\, \Ob_{12}.
\end{align}
Notably, unlike CLs, NGLs do not vanish in the case of no clustering (i.e., if $\Ob_{12} = 1$); instead, they reduce to the anti-$k_t$ case. Substituting for the irreducible eikonal amplitude, the NGLs contribution to the dijet mass cumulant is:
\begin{align}\label{eq:2loopKT:Sigma-NG}
	\S_{2, \ng}^{\ktt}(\rho) &= -2\,\asb^2\,\frac{L^2}{2!}\, \CF\CA\, \G_2^{\ktt}(R),
\end{align}
where the two loop NGLs coefficient is given by an expression analogous to that of the anti-$k_t$ case \eqref{eq:2loop:G2}, but with the integrand multiplied by the clustering condition factor $\Ob_{12}$:
\begin{align}\label{eq:2loopKT:G2}
	\G_2^{\ktt}(R) &= \frac{1}{2} \int_{1_\out} \int_{2_\inn} \A_{ab}^{12}\, \Ob_{12}.
\end{align}
The results as a function of jet-radius $R$ are shown in Fig. \ref{fig:F2G2} (B). We observe that the anti-$k_t$ NGLs coefficient is reduced by more than half due to $k_t$ clustering across the range of jet radii. Note that both NGLs coefficients tend to zero as $R \to \pi/2$, due to the decreasing phase space for harder emissions.

Comparing the full NGLs contribution (divided by $(\asb L)^2$) in the anti-$k_t$ algorithm, Eq. \eqref{eq:2loop:SigmaNG}, with that in the $k_t$ algorithm, which includes both CLs and NGLs, Eqs. \eqref{eq:2loopKT:Sigma-CL} and \eqref{eq:2loopKT:Sigma-NG}, we see in Fig. \ref{fig:F2G2} (C) that the anti-$k_t$ contribution is reduced by a factor of 2.5 for small values of $R$, with significantly higher reductions for $R>0.7$. This confirms previous findings \cite{Khelifa-Kerfa:2011quw, Kerfa:2012yae}, which suggest that using the $k_t$ algorithm with an optimal jet-radius can substantially mitigate NGLs effects on non-global observables. Consequently, the primary Sudakov form factor serves as a good approximation to the all-orders resummed result for such observables, thus avoiding the intricate task of resumming NGLs.

\begin{figure}
	\centering
	\includegraphics[scale=0.42]{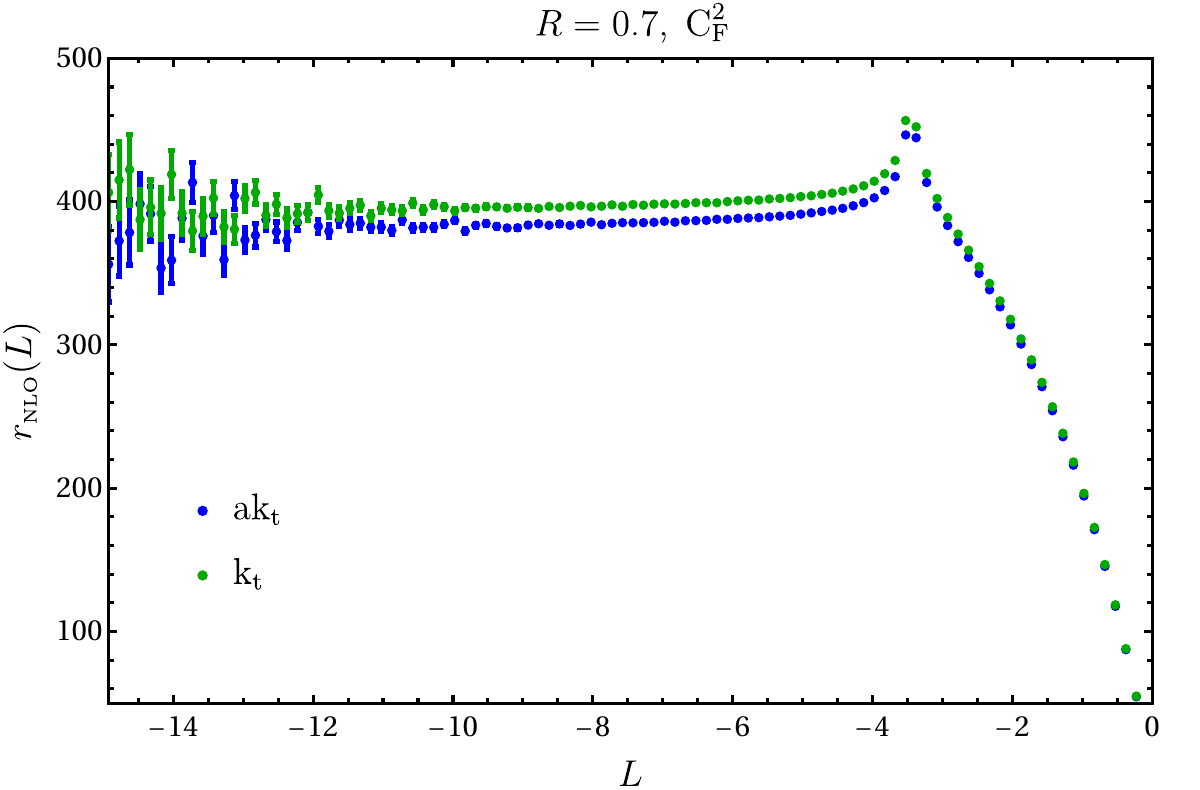} \vskip 0.5em
	\includegraphics[scale=0.42]{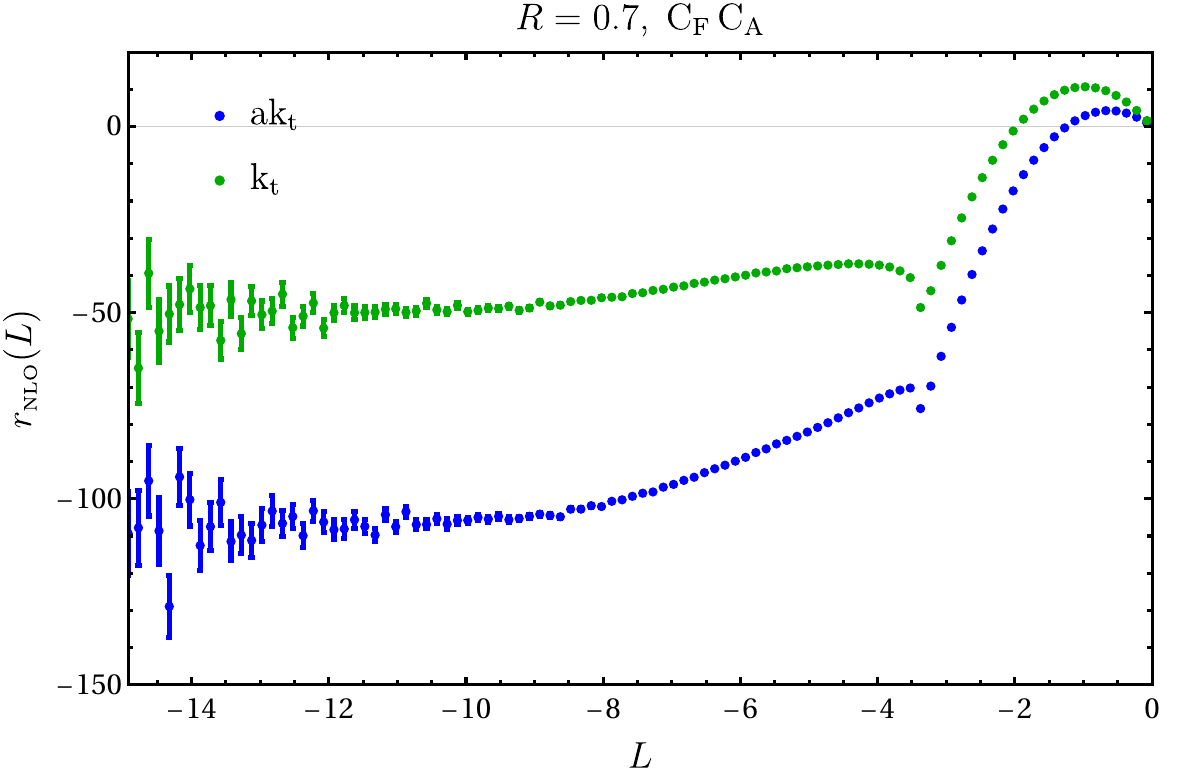} \vskip 0.5em
	\includegraphics[scale=0.42]{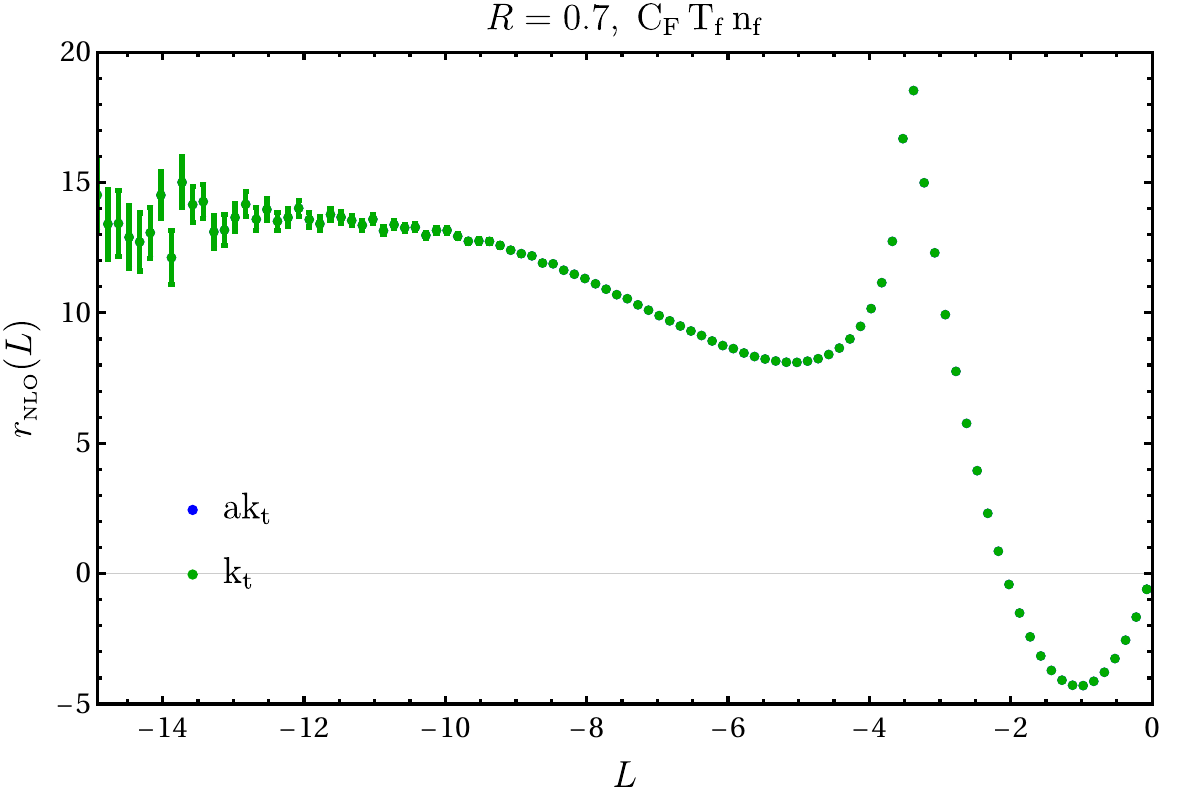}
	\caption{Comparisons between the analytical and \texttt{event2} differential distributions for the dijet mass observable at two loops for jet-radius $R = 0.7$.} \label{fig:2loop}
\end{figure}

The dijet mass distribution for the $k_t$ jet algorithm can thus be expressed at two loops as:
\begin{align}\label{eq:2loopKT:SigmaFull}
	\S_2^{\ktt}(\rho) = \frac{1}{2!} \left(\S_1^\P\right)^2 + \S_{2, \cl}^{\ktt} + \S_{2,\ng}^{\ktt}.
\end{align}
Finally, we compare the above result with the output from the fixed-order MC program \texttt{event2} for all three colour channels. Specifically, we plot in Fig. \ref{fig:2loop} the difference \eqref{eq:2loop:rNLO} for the choice $R = 0.7$. For large (negative) values of the logarithm $L$, this difference tends to a constant, indicating perfect cancellation of all logarithmic terms. In the $\CF \TF n_f$ channel, $k_t$ clustering has no effect at SL accuracy, which is why the two curves in Fig. \ref{fig:2loop} for this channel coincide. However, it does have an impact beyond SL, as shown in \cite{Kerfa:2012yae, Banfi:2021owj}.

In the following section, we present calculations of both CLs and NGLs at three loops in the $k_t$ algorithm. Notably, calculations of the NGLs component at this order have only recently appeared \cite{Benslama:2023gys} for the dijet azimuthal decorrelation observable. To the best of our knowledge, no similar calculations exist in the current literature.

\subsection{Three loops}
\label{sec:3loopKT}

\begin{figure}[t]
	\centering
	\includegraphics[scale=0.57]{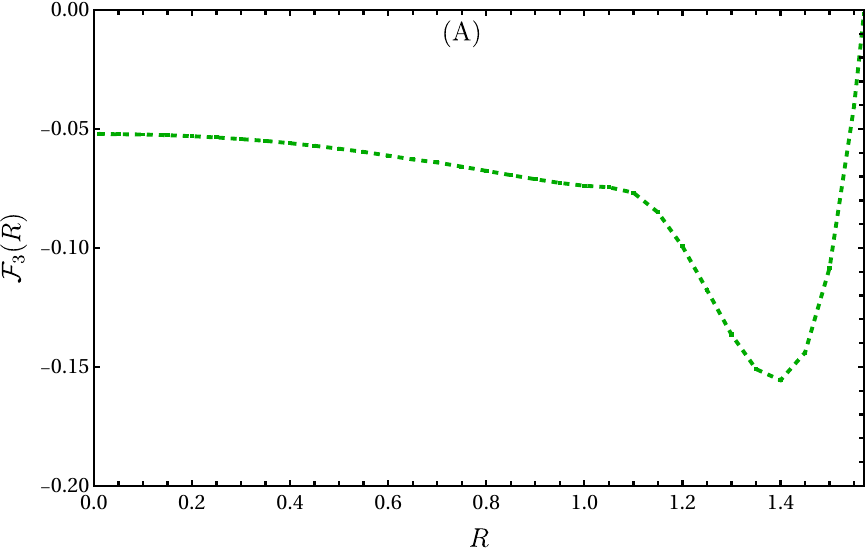} \vskip 0.5em
	\includegraphics[scale=0.57]{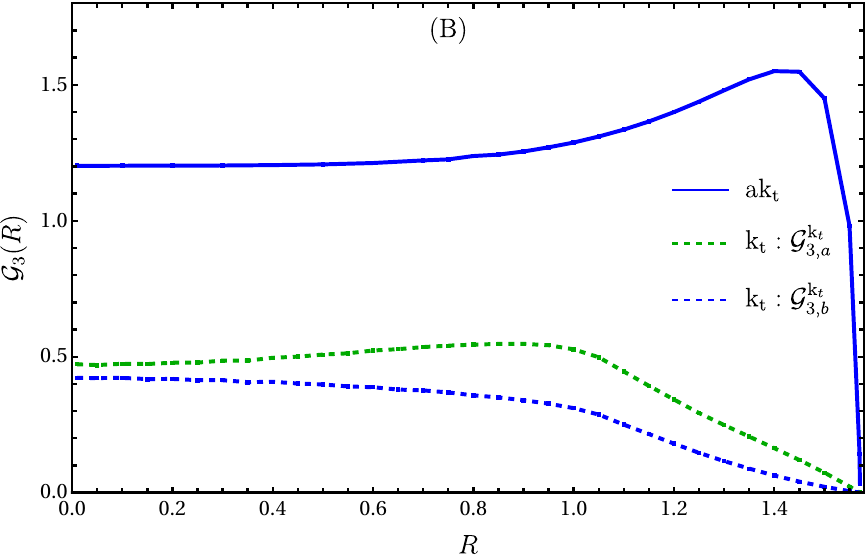} \vskip 0.5em
	\includegraphics[scale=0.54]{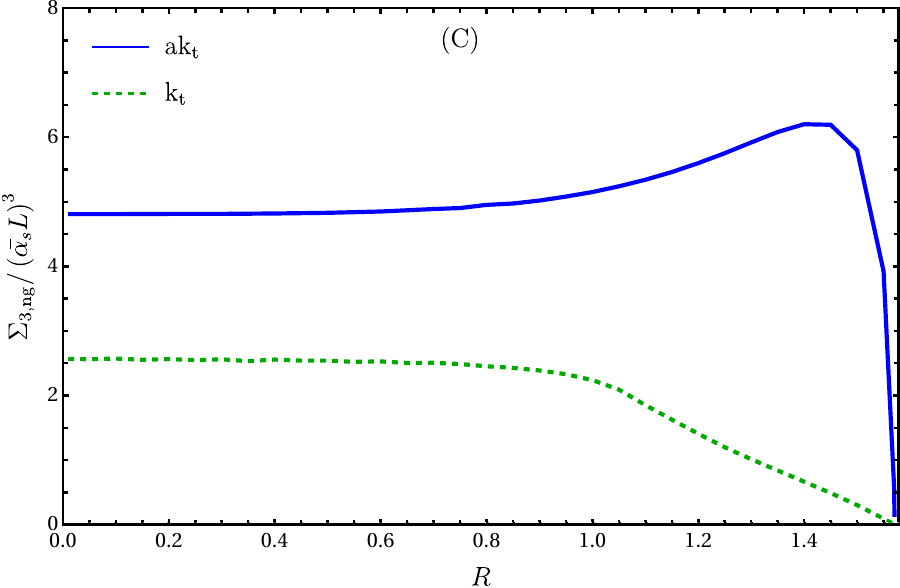}
	\caption{The three loop coefficients for CLs (A) and NGLs in both anti-$k_t$ (ak$_t$) and $k_t$ algorithms (B). A comparison between the NGLs and CLs+NGLs contributions to the dijet mass cumulant at three loops in anti-$k_t$ and $k_t$ algorithms, respectively (C).} \label{fig:F3G3}
\end{figure}

The sum over all possible gluon configurations for the emission of three soft, energy-ordered gluons is given by \cite{Khelifa-Kerfa:2024roc}:
\begin{align}\label{eq:3loopKT:uWX}
	\sum_X \Uh_3\W^\X_{123} &= -\Theta_1^\rho \Theta_2^\rho \Theta_3^\rho \Theta_3^\inn \Big[
	\W_{123}^{\V\V\R} + \Theta_1^\out\, \Ob_{13} \W_{123}^{\R\V\R} +
	\notag\\
	& + \Theta_2^\out\, \Ob_{23} \W_{123}^{\V\R\R} + \Theta_1^\out \left(\Theta_2^\out + \Theta_2^\inn \O_{12} \right) \W_{123}^{\R\R\R}
	\Big].
\end{align}
Note that in the absence of clustering ($\O_{ij} = 0$ for any pair of gluons $(ij)$), the above equation reduces to the anti-$k_t$ case, Eq. \eqref{eq:3loop:uWX}. Following the same steps outlined for the two loop calculations in the $k_t$ algorithm, we may express the full dijet mass distribution at three loops as:
\begin{align}\label{eq:3loopKT:SigmaFull}
	\S_3^{\ktt}(\rho) = \frac{1}{3!} \left(\S_1^\P\right)^3 + \S_1^\P \times \left(\S_{2, \cl}^{\ktt} + \S_{2,\ng}^{\ktt} \right) + \S_{3, \cl}^{\ktt} + \S_{3,\ng}^{\ktt}.
\end{align}
The first term on the rhs arises from the Sudakov form factor expansion, while the second term represents an interference term between the one-loop primary emissions contribution, Eq. \eqref{eq:1loop:Sigma-Final}, and the two loop CLs and NGLs contributions, Eqs. \eqref{eq:2loopKT:Sigma-CL} and \eqref{eq:2loopKT:Sigma-NG}, respectively. The third term is the new CLs contribution at this order, expressible in a form analogous to the two loop result \eqref{eq:2loopKT:Sigma-CL}:
\begin{align}\label{eq:3loopKT:Sigma-CL}
	\S_{3, \cl}^{\ktt}(\rho) = - 2\,\asb^3\,\frac{L^3}{3!}\,\CFcub\, \F_3(R),
\end{align}
where the three loop CLs coefficient is given by:
\begin{align}\label{eq:3loopKT:F3}
	\F_3(R) &= \Bigg[\int_{1_\out} \int_{2_\out} \int_{3_\inn} \,\O_{13} \O_{23}
	\notag\\
	&+ \int_{1_\out} \int_{2_\inn} \int_{3_\inn}\, \O_{12}
	\left(-1+\Ob_{13} \Ob_{23} \right)\Bigg] w_{ab}^1 w_{ab}^2 w_{ab}^3.
\end{align}
The result of this integration is plotted in Fig. \ref{fig:F3G3} (A). Notice that $\F_3$ is negative over the entire range of $R$, which makes the CLs contribution in Eq. \eqref{eq:3loopKT:Sigma-CL} positive. In the limit $R \to 0$, this result converges with that in \cite{Delenda:2012mm} for the single-jet-mass observable, $\F_3(R \sim 0) = -0.052$. The general features observed at two loops are retained at three loops, particularly the boundary nature of CLs and the fall-off to zero at large $R$ values. Additionally, the magnitude of $\F_3$ is smaller than that of $\F_2$ for all values of $R$, which ensures convergence of the CLs series.

The fourth term in Eq.~\eqref{eq:3loopKT:SigmaFull} represents the new NGLs contribution at three loops, and may be written as:
\begin{align}\label{eq:3loopKT:Sigma-NG}
	\S_{3,\ng}^{\ktt}(\rho) = + 2\,\asb^3\,\frac{L^3}{3!} \left[ \CFsq \CA\, \G_{3,a}^{\ktt} (R) + \CF \CAsq\, \G_{3,b}^{\ktt}(R) \right],
\end{align}
where the two parts of the NGLs coefficient are given by:
\begin{subequations}
	\begin{align}\label{eq:3loopKT:G3a}
		\G_{3,a}^{\ktt}(R) &= \frac{1}{2} \Bigg[
		\int_{1_\out} \int_{2_\out} \int_{3_\inn} \Big(\O_{13} \Ob_{23}\, w_{ab}^1 \A_{ab}^{23}
		+\Ob_{13} \O_{23} \times \notag\\
		& \qquad \times  \left[w_{ab}^2 \A_{ab}^{13} + w_{ab}^3 \A_{ab}^{12} \right] \Big)-
		\notag\\
		&   -\int_{1_\out} \int_{2_\inn} \int_{3_\inn} \Big( \O_{12} \Ob_{13} \Ob_{23} \left[ w_{ab}^1 \A_{ab}^{23} + w_{ab}^2 \A_{ab}^{13} \right] +
		\notag\\
		& + \left(\O_{13} +\O_{12} \left(-1+\Ob_{13} \Ob_{23}\right) \right) w_{ab}^3 \A_{ab}^{12}  \Big)  \Bigg],
	\end{align}
and
	\begin{align}\label{eq:3loopKT:G3b}
		\G_{3,b}^{\ktt}(R) &= \frac{1}{4} \Bigg[
		\int_{1_\out} \int_{2_\out} \int_{3_\inn} \Big( \Ob_{13} \O_{23} \A_{ab}^{12} \Ab_{ab}^{13}
		\notag\\
		&\qquad - \Ob_{13} \Ob_{23}\, \B_{ab}^{123} \Big) +
		\notag\\
		&+ \int_{1_\out} \int_{2_\inn} \int_{3_\inn} \Big( \Ob_{13} \left(1-\O_{12} \Ob_{23} \right) \A_{ab}^{12} \Ab_{ab}^{13} - \notag\\
		&\qquad - \O_{12} \Ob_{13} \Ob_{23}\, \B_{ab}^{123} \Big)
		\Bigg].
	\end{align}
\end{subequations}
The integration results for both terms are shown in Fig. \ref{fig:F3G3} (B). Several points are noteworthy regarding this contribution:
\begin{itemize}
	\item The $k_t$ clustering affects both the $\CFsq \CA$ and $\CF \CAsq$ colour channels, as well as the primary emissions channel ($\CFcub$). Therefore, clustering impacts all colour channels and types of emissions (primary, secondary, and their interferences). Note that in the anti-$k_t$ algorithm, there are no NGLs in the $\CFsq \CA$ colour channel.

	\item Both NGLs parts, $\G_{3,a}^{\ktt}$ and $\G_{3,b}^{\ktt}$, are smaller in magnitude than the anti-$k_t$ counterpart $\G_3^{\aktt}$, reduced by more than half across the range of $R$. The term $\G_{3,b}^{\ktt}$ (blue dashed curve in (B)) corresponds to the reduction in the anti-$k_t$ NGLs coefficient $\G_{3}^{\ktt}$ due to $k_t$ clustering, as they both share the same colour factor.

	\item The combined effect of both primary and secondary emissions in $k_t$ clustering again reduces the overall impact of anti-$k_t$ NGLs, as shown in Fig. \ref{fig:F3G3} (C). The reduction is approximately $50\%$, dominated by $\G_{3,b}^{\ktt}$ due to its large colour factor ($\CF \CAsq$).

	\item The clustering functions for both primary, $\Xi_{3,\cl}^{\ktt}$, and secondary, $\Xi_{3,\ng}^{\ktt}$, emissions can be inferred from Eq. \eqref{eq:3loopKT:uWX} upon substitution of the explicit eikonal amplitudes squared. However, we do not present them explicitly here as they are quite cumbersome. The CLs and NGLs coefficients in Eqs. \eqref{eq:3loopKT:F3}, \eqref{eq:3loopKT:G3a}, and \eqref{eq:3loopKT:G3b} include only parts of these clustering functions, excluding interference terms with lower orders.
\end{itemize}
The next section presents calculations with $k_t$ clustering at four loops. Unlike CLs, for which similar calculations exist in the literature \cite{Delenda:2006nf, Delenda:2012mm}, NGLs at this loop-order are unique to this work and have not previously appeared in the literature.

\subsection{Four loops}
\label{sec:4loopKT}

\begin{figure}[h!]
	\centering
	\includegraphics[scale=0.56]{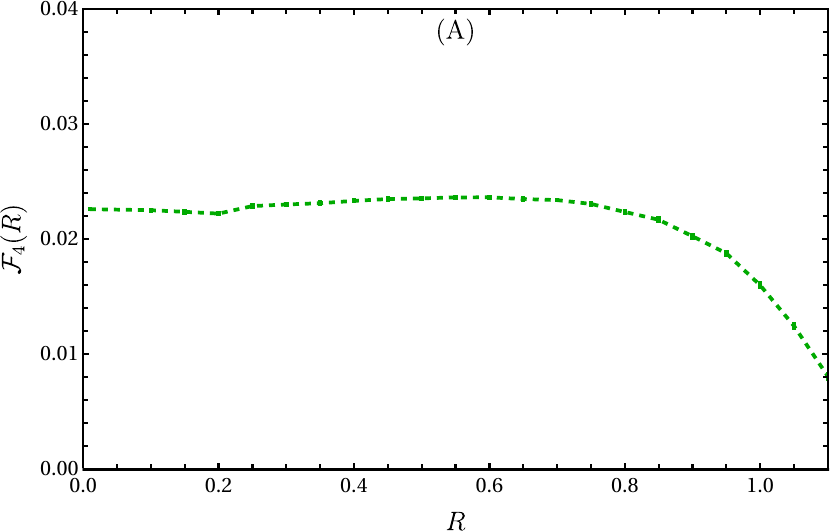}   \vskip 0.5em
	\includegraphics[scale=0.56]{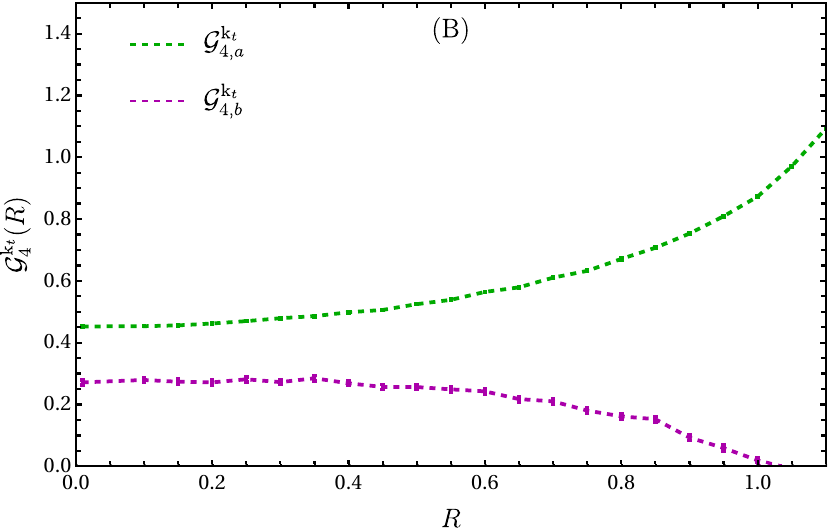} \vskip 0.5em
	\includegraphics[scale=0.56]{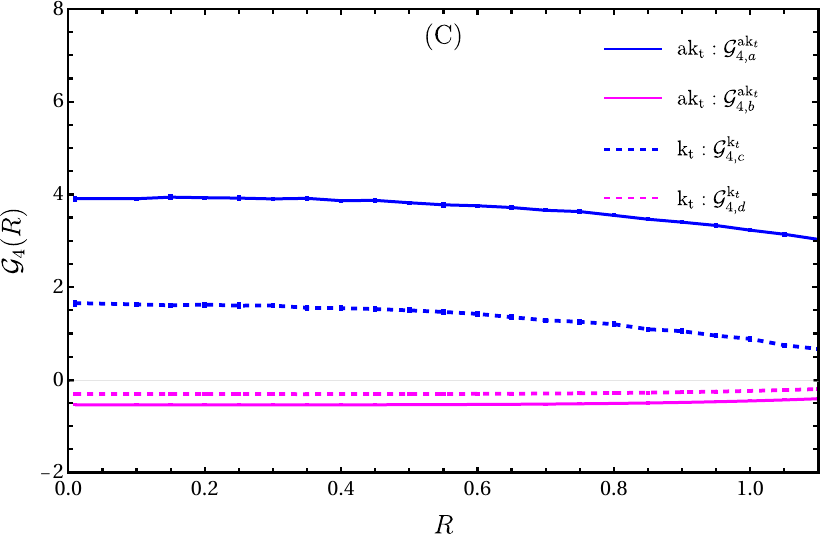} \vskip 0.5em
	\includegraphics[scale=0.4]{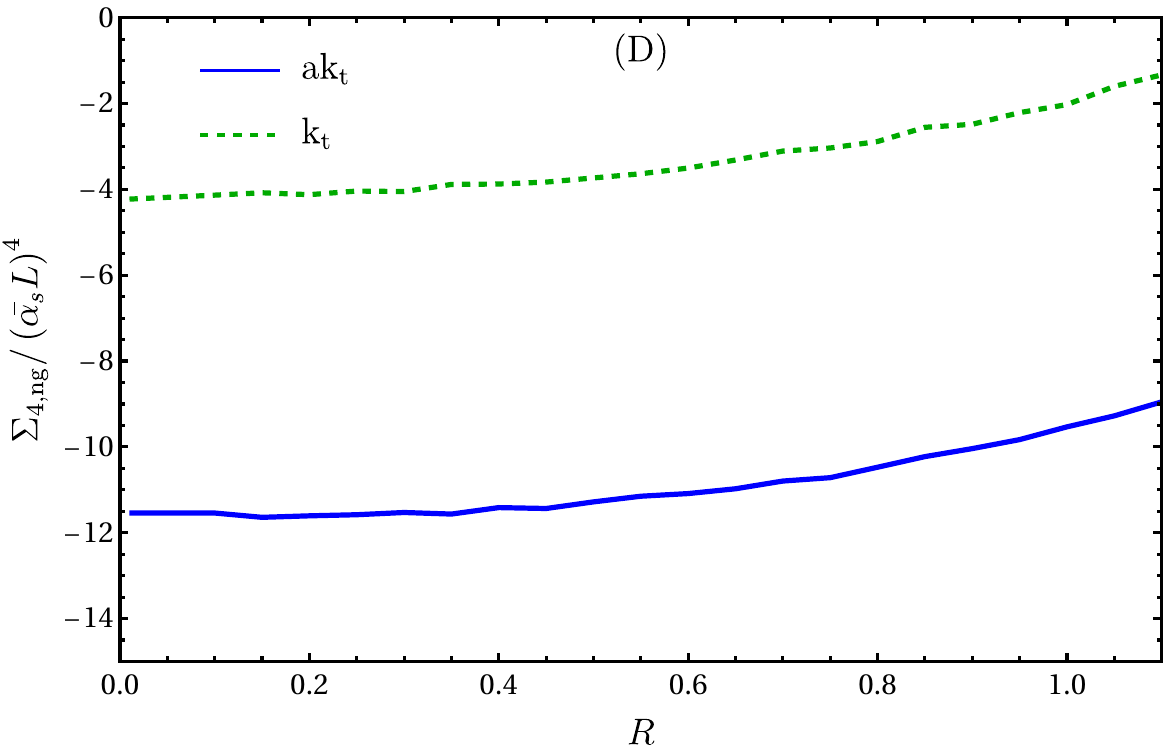}
	\caption{The four loop coefficients for CLs (A), NGLs unique to the $k_t$ algorithm (B), and NGLs present in both anti-$k_t$ (ak$_t$) and $k_t$ algorithms (C). Comparison between NGLs and CLs+NGLs contributions to the dijet mass cumulant at four loop order in the anti-$k_t$ and $k_t$ algorithms, respectively (D).} \label{fig:F4G4}
\end{figure}

The four loop calculations follow the same procedural steps as those established in previous lower loop orders. The sum over all possible gluon configurations of the squared eikonal amplitudes, in Eq. \eqref{eq:Def:Sigma-m} for $m=4$, is given in Eq. (20) of Ref.~\cite{Khelifa-Kerfa:2024roc}:
\begin{align}\label{eq:4loopKT:uWX}
	\sum_\X \Uh_4 \W_{1234}^{\X} &=-\Theta_1^\rho \Theta_2^\rho \Theta_3^\rho \Theta_4^\rho \Theta_4^\inn \Big[ \W_{1234}^{\V\V\V\R} + \Theta_1^\out \Ob_{14} \W_{1234}^{\R\V\V\R} +
	\notag\\
	&+ \Theta_2^\out \Ob_{24} \W_{1234}^{\V\R\V\R} + \Theta_3^\out \Ob_{34} \W_{1234}^{\V\V\R\R} +
	\notag\\
	&+ \Theta_1^\out \left(\Theta_2^\out + \Theta_2^\inn \O_{12} \right) \Ob_{14} \Ob_{24} \,\W_{1234}^{\R\R\V\R}
	\notag\\
	&+ \Theta_1^\out \left(\Theta_3^\out + \Theta_3^\inn \O_{13} \right) \Ob_{14} \Ob_{34} \,\W_{1234}^{\R\V\R\R}
	\notag\\
	&+ \Theta_2^\out \left(\Theta_3^\out + \Theta_3^\inn \O_{23} \right) \Ob_{24} \Ob_{34} \,\W_{1234}^{\V\R\R\R}
	\notag\\
	&+ \Theta_1^\out \left(\Theta_2^\out + \Theta_2^\inn \O_{12} \right) \times \notag\\
	&\times \left(\Theta_3^\out + \Theta_3^\inn \left[\O_{23} + \Ob_{23} \O_{13}\right]\right) \Ob_{14} \Ob_{24} \Ob_{34} \times
	\notag\\ &\,\W_{1234}^{\R\R\R\R}
	\Big].
\end{align}
Some of the important features and symmetry patterns of this formula are discussed in Ref.~\cite{Khelifa-Kerfa:2024roc}. Notice that, in the absence of clustering, the equation above reduces to Eq.~\eqref{eq:4loop:uWX} of the anti-$k_t$ case. Upon substituting the expressions for the various squared eikonal amplitudes, we obtain the four loop cumulant \eqref{eq:Def:Sigma-m}:
\begin{align}\label{eq:4loopKT:SigmaFull}
	\S_4^{\ktt}(\rho) &= \frac{1}{4!} \left(\S_1^\P\right)^4 + \S_1^\P \times \left(\S_{3,\cl}^{\ktt} + \S_{3,\ng}^{\ktt} \right) +
	\notag\\
	&+ \frac{1}{2!} \left(\S_1^\P \right)^2 \times \left(\S_{2,\cl}^{\ktt} + \S_{2,\ng}^{\ktt}\right) + \frac{1}{2!} \left( \S_{2,\cl}^{\ktt}\right)^2 +
	\notag\\
	&+ \frac{1}{2!} \left( \S_{2,\ng}^{\ktt}\right)^2 + \S_{4,\cl}^{\ktt} + \S_{4,\ng}^{\ktt}.
\end{align}
The last two terms represent the new CLs and NGLs contributions at this order. The CLs term, $\S_{4,\cl}^{\ktt}$, may be cast, as usual, in the form:
\begin{align}\label{eq:4loopKT:Sigma-CL}
	\S_{4,\cl}^{\ktt}(\rho) = +2\,\asb^4\, \frac{L^4}{4!}\,\CFfour\, \F_4(R),
\end{align}
where the CLs four loop coefficient has the explicit expression:
\begin{align}\label{eq:4loopKT:F4}
	\F_4(R) &= \Bigg[
	\int_{1_\out} \int_{2_\out} \int_{3_\out} \int_{4_\inn} \O_{14} \O_{24} \O_{34}
	\notag\\
	&+\int_{1_\out} \int_{2_\out} \int_{3_\inn} \int_{4_\inn} \Big[\O_{13} \O_{24} \left(-1+\Ob_{14} \Ob_{34} \right) + \O_{23} \times
	\notag\\& \times \left(-\O_{13} -\O_{14} +\Ob_{24} \Ob_{34} \left(1-\Ob_{13} \Ob_{14}\right) \right) \Big]
	\notag\\
	&+\int_{1_\out} \int_{2_\inn} \int_{3_\out} \int_{4_\inn} \O_{12} \O_{34} \left(-1+\Ob_{14} \Ob_{24}\right)
	\notag\\
	&+ \int_{1_\out} \int_{2_\inn} \int_{3_\inn} \int_{4_\inn} \O_{12} \left(-1+\Ob_{13} \Ob_{23}\right) \times
	\notag\\
	&\times \left(-1+\Ob_{14} \Ob_{24} \Ob_{34}\right) \Bigg] w_{ab}^1 w_{ab}^2 w_{ab}^3 w_{ab}^4.
\end{align}
The results of the integration are plotted in Fig.~\ref{fig:F4G4} (A). For small values of the jet radius $R$, one recovers the value reported in \cite{Delenda:2012mm}, specifically $\F_4(R \sim 0) = 0.0226$.

Similarly, the NGLs term at this order can be expressed as:
\begin{align}\label{eq:4loopKT:G4}
	\S_{4,\ng}^{\ktt}(\rho) &= - 2\,\asb^2\,\frac{L^4}{4!} \Big[-\CFcub \CA\, \G_{4,a}^{\ktt} - \CFsq \CAsq\, \G_{4,b}^{\ktt} +
	\notag\\
	&\qquad +  \CF \CAcub\, \G_{4,c}^{\ktt} +  \CF \CAsq \left(\CA - 2\CF\right) \G_{4,d}^{\ktt} \Big],
\end{align}
where the expressions of the various NGLs terms are complex, and we refrain from showing them explicitly. The final integration results are displayed in Fig.~\ref{fig:F4G4} ((B) and (C)) as a function of the jet radius $R$. Notably, both $\CF \CAcub$ and finite-$N_c$ coefficients, $\G_{4,c}^{\ktt}$ and $\G_{4,d}^{\ktt}$ in the equation above, are negative, and a minus sign has been extracted from the former, $\G_{4,c}^{\ktt}$, to maintain analogy with the anti-$k_t$ case ($\S_{4,\ng}^{\aktt}$ in Eq.~\eqref{eq:4loop:SigmaNG-B}). All characteristics observed at lower orders persist at four loop order, including that $k_t$ clustering affects all colour channels and that the dominant contribution arises from the term associated with the largest colour factor ($\CF \CAcub$).
The NGLs reduction due to clustering at this order exceeds $50\%$ across all values of the jet-radius, as demonstrated in Fig.~\ref{fig:F4G4} (D).

\subsection{Exponentiation}

The calculations of the dijet mass distribution in both anti-$k_t$ and $k_t$ jet algorithms up to four loops, Eqs.~\eqref{eq:2loop:Sigma}, \eqref{eq:3loop:Sigma-A}, \eqref{eq:4loop:Sigma}, \eqref{eq:2loopKT:SigmaFull}, \eqref{eq:3loopKT:SigmaFull}, and \eqref{eq:4loopKT:SigmaFull}, clearly indicate a pattern of exponentiation for both CLs and NGLs.
For instance, the jet-mass cumulant at four loops in the anti-$k_t$ algorithm, Eq.~\eqref{eq:4loop:Sigma}, corresponds to the $(\asb L)^4$ term in the expansion of the following product of exponentials:
\begin{align}\label{eq:Exp:AKT}
	e^{\S_1^\P} \times e^{\S^{\aktt}_{2,\ng} + \S^{\aktt}_{3, \ng} + \S^{\aktt}_{4, \ng}}
	&= \Big[ 1 + \S_1^\P + \frac{1}{2!} \left(\S_1^\P\right)^2 + \frac{1}{3!} \left(\S_1^\P\right)^3 + \notag\\
	&+ \frac{1}{4!} \left(\S_1^\P\right)^4 + \cdots \Big] \Big[ 1 + \S^{\aktt}_{2,\ng} + \S^{\aktt}_{3, \ng} + \notag\\
	&+ \S^{\aktt}_{4, \ng} + \frac{1}{2!} \left(\S^{\aktt}_{2,\ng}\right)^2 + \cdots \Big],
\end{align}
where $\S_1^\P \propto \asb L$, $\S^{\aktt}_{j, \ng} \propto (\asb L)^j$, and the ellipsis denotes terms beyond $(\asb L)^4$. Similarly, the jet-mass cumulant at four loops in the $k_t$ algorithm, Eq.~\eqref{eq:4loopKT:SigmaFull}, is given by the $(\asb L)^4$ term in the expansion of:
\begin{align}\label{eq:Exp:KT}
	e^{\S_1^\P} \times e^{\S^{\aktt}_{2,\ng} + \S^{\aktt}_{3, \ng} + \S^{\aktt}_{4, \ng}} \times e^{\S^{\ktt}_{2, \cl} + \S^{\ktt}_{3, \cl} + \S^{\ktt}_{4, \cl}},
\end{align}
where $\S^{\ktt}_{j, \cl} \propto (\asb L)^j$.
The origin of the exponentiation of the jet-mass distribution lies in the exponentiation pattern of the squared eikonal emission amplitudes, as explained in detail in Ref.~\cite{Delenda:2015tbo}. For the $k_t$ algorithm, it also stems from the exponentiation of the phase-space clustering functions, as highlighted in Ref.~\cite{Khelifa-Kerfa:2024roc}. Further details can be found in Refs.~\cite{Delenda:2012mm, Khelifa-Kerfa:2015mma}.

By substituting the explicit expressions for the various NGLs and CLs jet-mass cumulants and extending the exponentiation in Eqs.~\eqref{eq:Exp:AKT} and \eqref{eq:Exp:KT} to $n$ loops, we obtain:
\begin{subequations}
	\begin{align}
		\cS^{\mathrm{\scriptscriptstyle JA}}(\rho) &= \exp\left[ -2 \sum_{n\geq 2} \frac{1}{n!}\, \cS_{n}^{\mathrm{\scriptscriptstyle JA}}(R)\, \left(-\asb\, L\right)^n \right], \label{eq:FormFactor-NGLs}
		\\
		\C(\rho) &= \exp\left[ 2\,\sum_{n \geq 2} \frac{1}{n!} \, \F_n(R)\, \left(-\CF\,\asb\,L\right)^n \right], \label{eq:FormFactor-CLs}
	\end{align}
\end{subequations}
where the superscript $\mathrm{JA}$ refers to a given jet algorithm, i.e., anti-$k_t$ or $k_t$, and hence $\cS_n^{\aktt}$, for the anti-$k_t$, is given by:
\begin{subequations}
	\begin{align}
		& \cS_2^{\aktt} = \CF \CA \,\G_{2}^{\aktt}, \qquad \cS_3^{\aktt} = \CF \CAsq \,\G_{3}^{\aktt},
		\notag\\
		& \cS_4^{\aktt} = \CF \CAcub \,\G_{4,a}^{\aktt} + \CF\CAsq \left(\CA - 2\CF\right) \G_{4,b}^{\aktt},
	\end{align}
	and for \kt:
	\begin{align}
		& \cS_2^{\ktt} = \CF \CA \,\G_{2}^{\ktt}, \qquad
		\cS_3^{\ktt} = \CFsq \CA\,\G_{3,a}^{\ktt} + \CF \CAsq \,\G_{3,b}^{\aktt},
		\notag\\
		& \cS_4^{\ktt} = -\CFcub \CA \,\G_{4,a}^{\ktt} - \CFsq \CAsq\,\G_{4,b}^{\ktt} + \CF \CAcub \,\G_{4,c}^{\aktt} +
		\notag\\
		&\qquad + \CF\CAsq \left(\CA - 2\CF \right) \G_{4,d}^{\ktt}.
	\end{align}
\end{subequations}
Due to the alternating signs in both the CLs and NGLs series (for both jet algorithms), and to assess their convergence and approximation to the all-orders result, we shall compare the exponentials in Eqs.~\eqref{eq:FormFactor-NGLs} and \eqref{eq:FormFactor-CLs} to the all-orders numerical output of the MC code of \cite{Dasgupta:2001sh} in Section~\ref{sec:MC}. Although this code computes NGLs in the large-$N_c$ limit, it remains the only available code that implements $k_t$ clustering and accommodates various non-global observables. Finite-$N_c$ MC codes, such as those in \cite{Hatta:2013iba, Hagiwara:2015bia}, are restricted to specific observables (excluding the dijet mass) and lack $k_t$ clustering.

Before conducting these comparisons, we shall first outline, in the next section, the all-orders resummed formula for the dijet mass distribution, incorporating both CLs and NGLs contributions.

\section{All-orders resummation}
\label{sec:Resummation}

The resummation of large logarithms appearing in the distribution of the dijet mass non-global observable can be cast in the following factorised formula \cite{Dasgupta:2001sh, Khelifa-Kerfa:2011quw}:
\begin{align}\label{eq:AllOrders:ResumFormFactor-Full}
	\S^{\mathrm{\scriptscriptstyle JA}}(\rho) = \S^\P(\rho)\, \cS^{\mathrm{\scriptscriptstyle JA}}(t)\, \C(t),
\end{align}
where $\S^\P$ is the well-known Sudakov form factor that resums primary emissions and takes the usual standard form \cite{Catani:1992ua}:
\begin{align}\label{eq:AllOrders:ResumFormFactor-Primary}
	\S^\P(\rho) = \frac{ \exp\left(-\mathcal{R} -\gamma_E \, \mathcal{R}' \right)  }{ \Gamma \left(1 + \mathcal{R}'\right) },
\end{align}
where $\gamma_E = 0.577$ is the Euler–Mascheroni constant, $\mathcal{R}$, known as the global radiator, is a function of $\rho$ and $R^2$ that represents the resummation of large global logarithms to NLL accuracy, and $\mathcal{R}'$ is its derivative with respect to the logarithm $L'= \log(R^2/\rho)$. The full expression of this radiator and its derivative can be found in Refs.~\cite{Banfi:2010pa, Kerfa:2012yae}.

The NGLs form factor, $\cS^{\mathrm{\scriptscriptstyle JA}}(t)$, can be parametrised by the following exponential, initially proposed in \cite{Dasgupta:2001sh}:
\begin{align}\label{eq:AllOrders:ResumFormFactor-NGLs}
	\cS_{\mathrm{\scriptscriptstyle MC}}^{\mathrm{\scriptscriptstyle JA}}(t) = \exp\left[ -\CF\CA\, \G_2^{\mathrm{\scriptscriptstyle JA}}(R)\, \left( \frac{1 + (a t)^2}{ 1 + (b t)^c} \right)\, t^2 \right],
\end{align}
where $a$, $b$, and $c$ are fitting parameters to the MC output, $\G_2^{\mathrm{\scriptscriptstyle JA}}$ is the two loop NGLs coefficient for the jet algorithm $\mathrm{JA}$ (anti-$k_t$ or $k_t$), and $t$ is the evolution variable, given at SL accuracy by:
\begin{align}
	t = \frac{1}{2\pi} \int_{\frac{\rho}{R^2}}^{1} \frac{\d x}{x}\, \as\left(x\,Q/2\right)
	= - \frac{1}{4 \pi \beta_0} \ln \left[ 1 - 2\as \beta_0 \, L'\right],
\end{align}
with $\as\left(Q R/2\right)$ and $L'$ defined as above. The CLs form factor assumes an analogous form:
\begin{align}\label{eq:AllOrders:ResumFormFactor-CLs}
	\C_{\mathrm{\scriptscriptstyle MC}}(t) = \exp\left[\CFsq\, \F_2(R)\, \left( \frac{1 + (a t)^2}{ 1 + (b t)^c} \right)\, t^2 \right],
\end{align}
where $\F_2$ is the two loop CLs coefficient in the $k_t$ algorithm. The fitting parameters depend on the jet radius, the jet algorithm, and differ for NGLs and CLs resummations. In Tables~\ref{tab:FittingParams-NGLs} and \ref{tab:FittingParams-CLs}, we show the values of the fitting parameters used for the NGLs and CLs parametrisations, respectively.
\begin{table}[h!]
	\centering
	\begin{tabular}{>{\centering\arraybackslash}m{2cm} >{\centering\arraybackslash}m{0.7cm} >{\centering\arraybackslash}m{0.7cm} >{\centering\arraybackslash}m{0.7cm} >{\centering\arraybackslash}m{0.7cm} >{\centering\arraybackslash}m{0.7cm}}
		\toprule
		\textbf{Jet algorithm} & $R$   & $\G_2$ & $a$ & $b$ & $c$ \\
		\midrule
		anti-$k_t$ 			   & $0.7$ & $6.51$ & $0.02\CA$ & $0.58\CA$ & $1.48$ \\
		& $1.0$ & $6.21$ & $0.01\CA$ & $0.92\CA$ & $1.81$ \\
		$k_t$        			   & $0.7$ & $2.54$ & $0.96\CA$ & $0.44\CA$ & $0.14$ \\
		& $1.0$ & $1.87$ & $0.91\CA$ & $2.99\CA$ & $0.15$ \\
		\bottomrule
	\end{tabular}
	\caption{Values of the fitting parameters for the parametrisation of the NGLs resummed form factor.} \label{tab:FittingParams-NGLs}
\end{table}
%
\begin{table}[h!]
	\centering
	\begin{tabular}{>{\centering\arraybackslash}m{2cm} >{\centering\arraybackslash}m{0.7cm} >{\centering\arraybackslash}m{0.7cm} >{\centering\arraybackslash}m{0.7cm} >{\centering\arraybackslash}m{0.7cm} >{\centering\arraybackslash}m{0.7cm}}
		\toprule
		\textbf{Jet algorithm} & $R$   & $\F_2$ & $a$ & $b$ & $c$ \\
		\midrule
		$k_t$ 		 		   & $0.7$ & $0.96$ & $0.57\CA$ & $0.72\CA$ & $1.62$ \\
		& $1.0$ & $1.31$ & $0.19\CA$ & $0.25\CA$ & $0.42$ \\

		\bottomrule
	\end{tabular}
	\caption{Values of the fitting parameters for the parametrisation of the CLs resummed form factor.} \label{tab:FittingParams-CLs}
\end{table}

\begin{figure}[h!]
	\centering
	\includegraphics[scale=0.57]{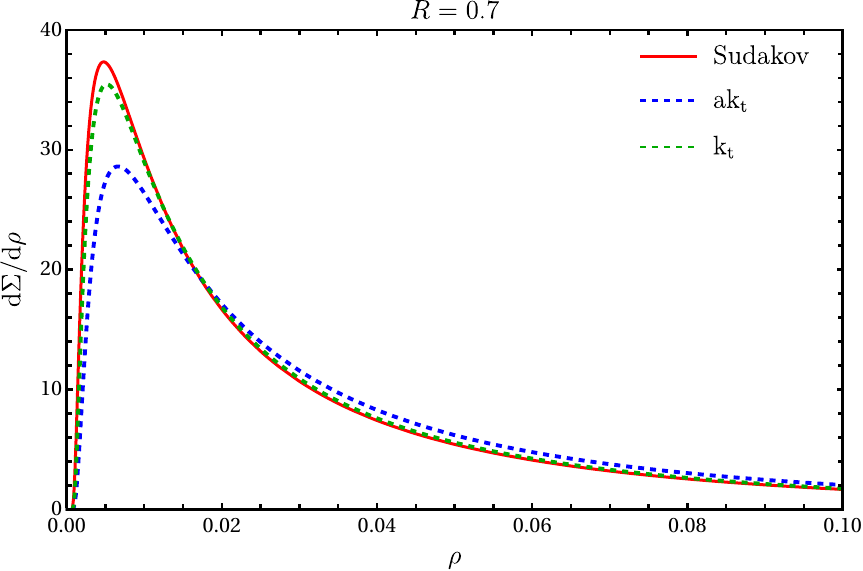} \vskip 0.5em
	\includegraphics[scale=0.57]{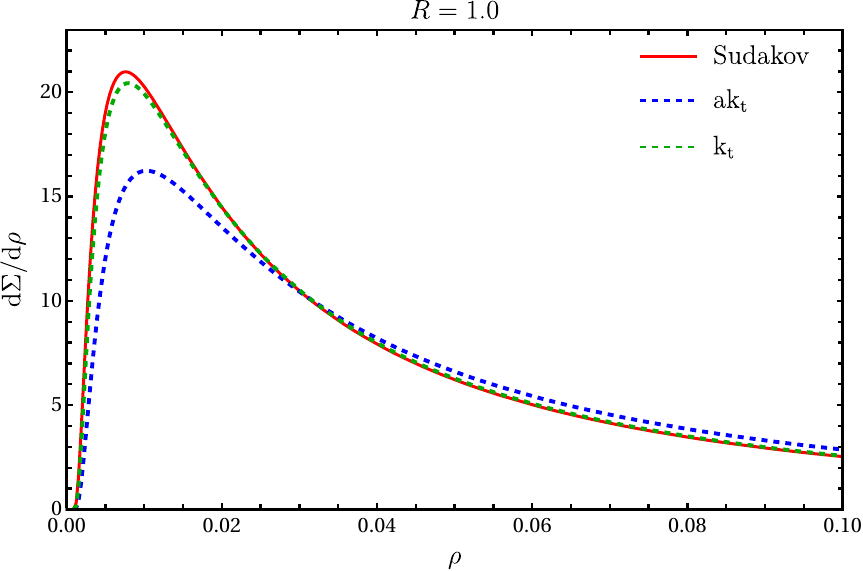}
	\caption{The resummed distribution for the dijet mass observable \eqref{eq:AllOrders:ResumFormFactor-Full} with and without the inclusion of NGLs and CLs form factors for both anti-$k_t$ and $k_t$ jet algorithms at $R = 0.7$ and $R=1.0$. } \label{fig:Resum}
\end{figure}

The resummed distribution \eqref{eq:AllOrders:ResumFormFactor-Full} is plotted in Fig.~\ref{fig:Resum} for both anti-$k_t$ and $k_t$ jet algorithms. For $R = 0.7$, the peak of the Sudakov distribution is shifted towards higher values of $\rho$ in the anti-$k_t$ algorithm and is reduced by about $24\%$. In the $k_t$ algorithm, the position of this peak remains unchanged and is reduced by only $5\%$. For $R = 1.0$, the same observations hold, with reductions of about $22\%$ and $2\%$ for anti-$k_t$ and $k_t$ algorithms, respectively. Larger values of $R$ would result in even smaller reductions, particularly for the $k_t$ algorithm.
These results reinforce previous conclusions \cite{Khelifa-Kerfa:2011quw, Kerfa:2012yae} that selecting the $k_t$ algorithm over the anti-$k_t$ algorithm has the advantage of minimising the complex effect of NGLs, thereby making the Sudakov primary resummation more reliable.
%

\subsection{Comparisons of analytical and MC results}
\label{sec:MC}

\begin{figure}[h!]
	\centering
	\includegraphics[scale=0.41]{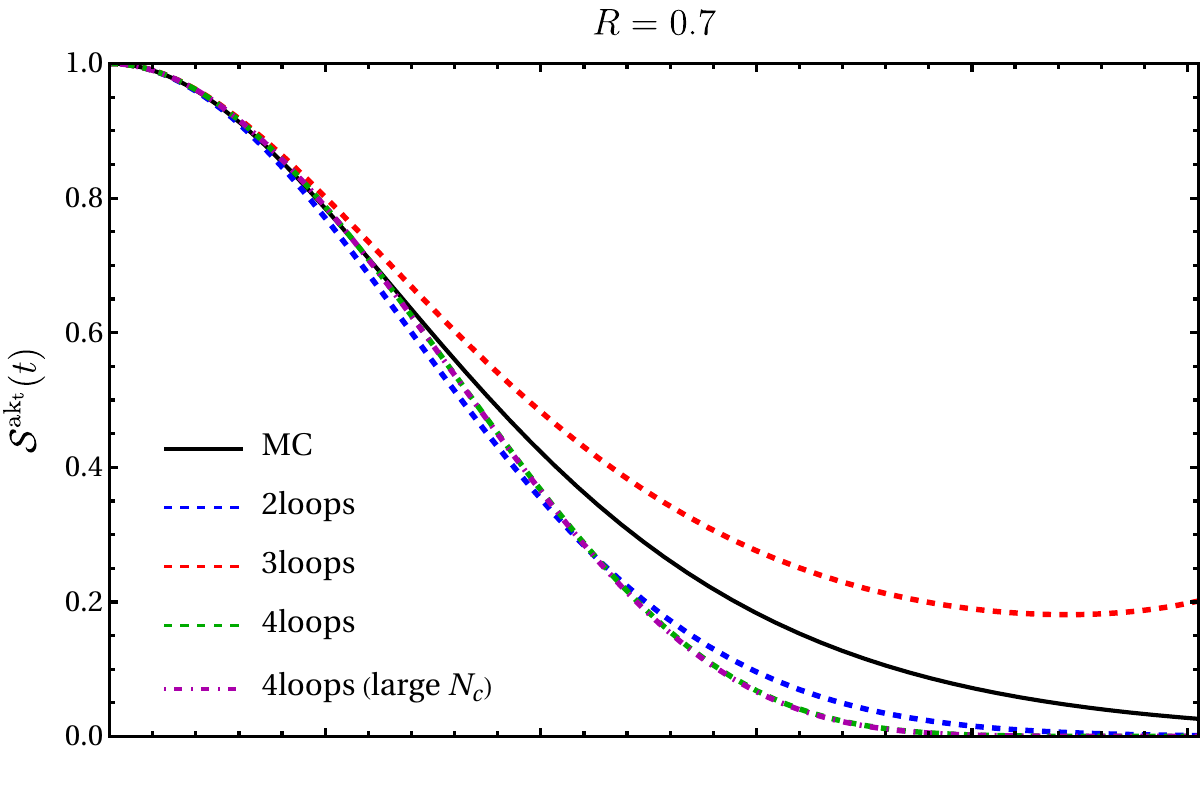} \vskip -2em
	\includegraphics[scale=0.41]{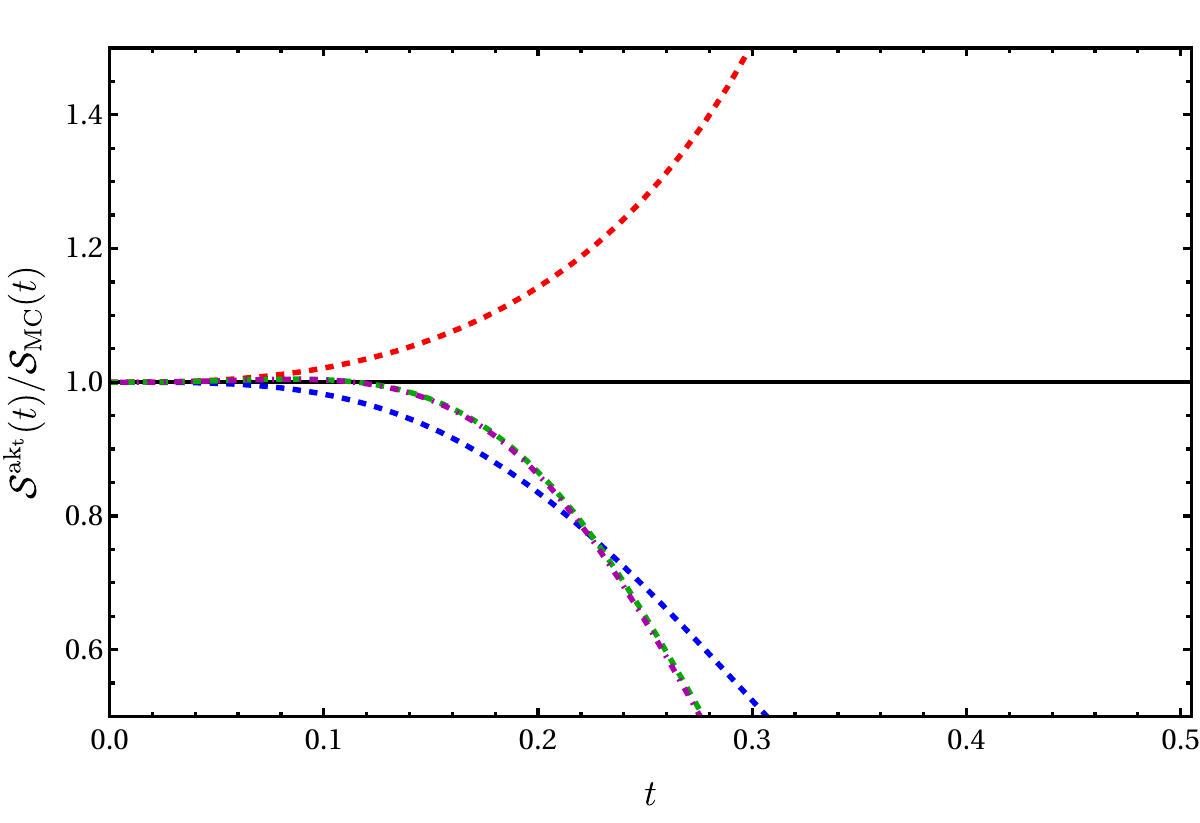}
	\caption{Comparison between the exponential of fixed-order NGLs results \eqref{eq:FormFactor-NGLs}, both at finite-$N_c$ and large-$N_c$, and the output of the MC code from \cite{Dasgupta:2001sh} for the anti-$k_t$ algorithm with $R = 0.7$.} \label{fig:St-R7-akt}
\end{figure}

\begin{figure}[h!]
	\centering
	\includegraphics[scale=0.40]{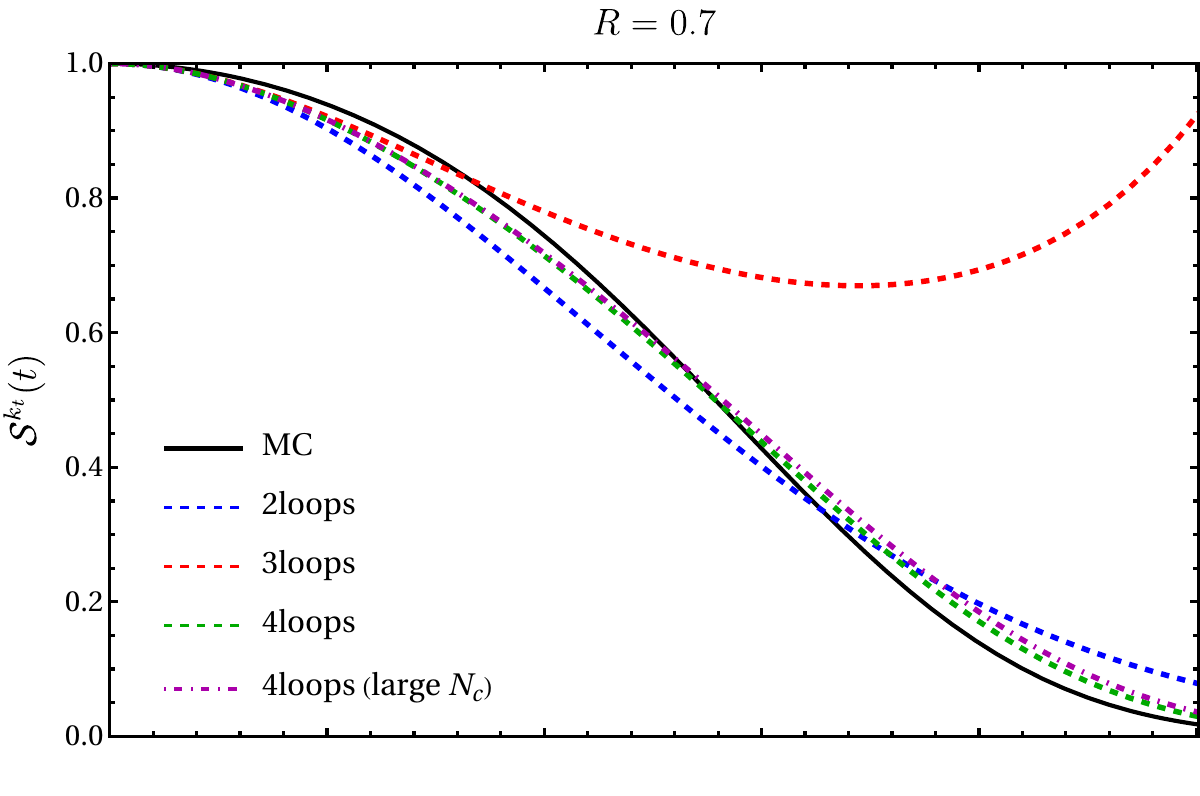} \vskip -2em
	\includegraphics[scale=0.41]{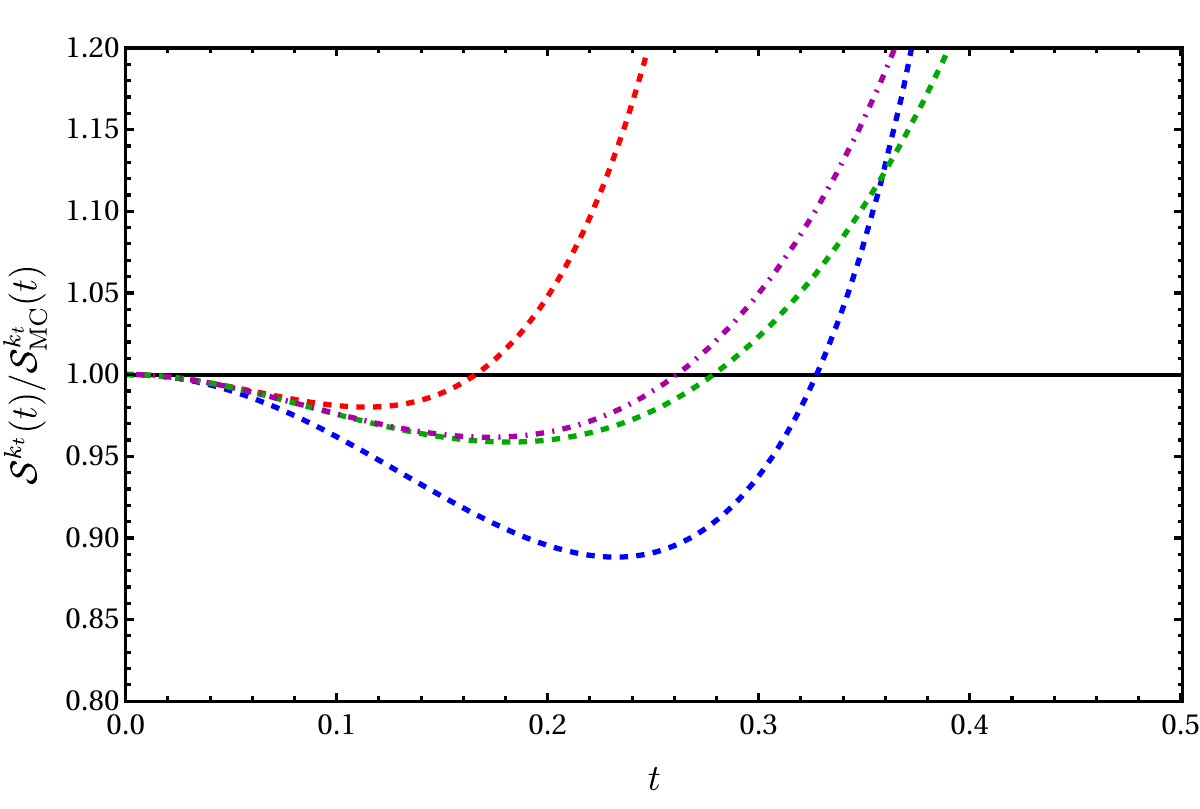}
	\caption{Comparison between the exponential of fixed-order NGLs results \eqref{eq:FormFactor-NGLs}, both at finite-$N_c$ and large-$N_c$, and the output of the MC code from \cite{Dasgupta:2001sh} for the $k_t$ algorithm with $R = 0.7$.} \label{fig:St-R7-kt}
\end{figure}

\begin{figure}[h!]
	\centering
	\includegraphics[scale=0.57]{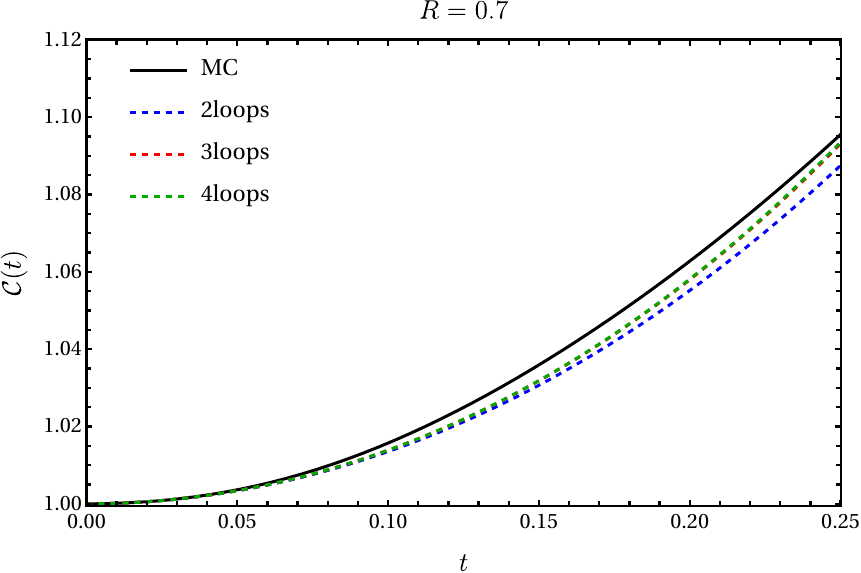} \vskip 0.5em
	\caption{Comparison between the exponential of fixed-order CLs results \eqref{eq:FormFactor-CLs} and the output of the MC code from \cite{Dasgupta:2001sh} for the $k_t$ algorithm with $R = 0.7$.} \label{fig:Ct-R7-kt}
\end{figure}

In this section, we compare the analytical exponentiation of the fixed-order calculations of NGLs and CLs terms, Eqs.~\eqref{eq:FormFactor-NGLs} and \eqref{eq:FormFactor-CLs}, with the numerical distributions obtained from the MC of \cite{Dasgupta:2001sh}, represented by the parametrisations \eqref{eq:AllOrders:ResumFormFactor-NGLs} and \eqref{eq:AllOrders:ResumFormFactor-CLs}. At fixed-order, $t = \asb L'/2$, so we replace $\asb L'$ in Eqs.~\eqref{eq:FormFactor-NGLs} and \eqref{eq:FormFactor-CLs} by $2t$. Furthermore, the legends ``2loops", ``3loops" and ``4loops" in Figs.~\ref{fig:St-R7-akt}, \ref{fig:St-R7-kt}, and \ref{fig:Ct-R7-kt} indicate truncating the series in the exponents of these equations at two, three, and four loops, respectively.

For the anti-$k_t$ jet algorithm, the two loop NGLs exponential approximates the all-orders NGLs distribution quite well over the entire range of $t$. The case is less accurate for the $k_t$ algorithm. Adding contributions from higher loop-orders improves the approximation at small (but phenomenologically important) values of $t$. This is particularly true when up to four loop terms are included in the exponent of the analytical expressions.

The curves labelled ``4 loops (large-$N_c$)'' in Figures~\ref{fig:St-R7-akt} and \ref{fig:St-R7-kt} correspond to the scenario where the finite-$N_c$ terms at four loops are set to zero. Specifically, these terms are proportional to the colour factor $\left(\CA - 2\CF\right)$, represented by $\G_{4,b}^{\aktt}$ and $\G_{4,d}^{\ktt}$ for the anti-$k_t$ and $k_t$ algorithms, respectively. As is evident from the preceding calculations, this is the only contribution that differentiates finite-$N_c$ from large-$N_c$ results up to four loop. Notably, the two and three loop distributions are identical between the finite-$N_c$ and large-$N_c$ cases.
The plots suggest that the finite-$N_c$ corrections are relatively minor, especially for the anti-$k_t$ algorithm, where the curves are indistinguishable over the entire range of $t$. For $R=0.7$, the finite-$N_c$ corrections result in a reduction of the four loop coefficient by approximately $1.6\%$ compared to the large-$N_c$ result. This reduction becomes negligible when multiplied by $1/4!$ and exponentiated (as in Eq. \eqref{eq:FormFactor-NGLs}). Moreover, the $k_t$ algorithm exhibits a larger reduction of about $4.4\%$, which manifests as a subtle yet discernible deviation between the curves in Fig.~\ref{fig:St-R7-kt}, particularly at higher values of $t$.

The minor effect of finite-$N_c$ corrections in both algorithms is attributed to the smallness of their corresponding colour factor, $\CF \CAsq (\CA - 2\CF) = \CF \CA$, compared to $\CF \CAcub$ of the large-$N_c$ term (i.e., a factor of $\CAsq$ smaller). This observation has been clearly stated in previous sections. The slightly larger effect of finite-$N_c$ observed for the $k_t$ algorithm arises from the fact that the four loop NGLs term proportional to $\CF \CAcub$, namely $\G^{\aktt}_{4,a}$, experiences a greater reduction due to clustering compared to the finite-$N_c$ term, $\G^{\aktt}_{4,b}$, as is evident in Fig.~\ref{fig:F4G4}(C).
This is in addition to the fact that the finite-$N_c$ coefficient is already small, even for the anti-$k_t$ algorithm ($\G^{\aktt}_{4,b} \sim -\zeta_4/2$ over the whole range of $R$), and hence its reduction due to clustering does not have a significant impact when compared to that of the $\CF \CAcub$ coefficient.

It is worth noting that for comparisons of the CLs distributions (Fig.~\ref{fig:Ct-R7-kt}), we used the parametrisation adopted in the MC code \cite{Dasgupta:2001sh} and implemented in our previous paper \cite{Delenda:2012mm} (Eq.~(A.8)). In this parametrisation, the numerical values of the CLs coefficients match those found here for small values of $R$ (up to about $R = 0.5$) but are generally slightly lower for larger values.
The CL series appears to converge to the all-orders result faster than the NGLs series. This is evident from the decreasing (absolute) values of the CL coefficients at higher loop orders, as well as from the ``4loops'' curve in Fig.~\ref{fig:Ct-R7-kt}. Notably, the CL form factor contributes, for $R=0.7$, up to a maximum of $10\%$ of the resummed primary emissions for values of $t$ up to $0.25$. Smaller percentage contributions are observed for smaller values of $t$.

\section{Conclusion}
\label{sec:Conclusion}

In this paper, we have investigated the fixed-order and all-orders resummed distribution of the normalised invariant mass (squared) observable for dijets produced at threshold in $e^+ e^-$ annihilation processes. This observable belongs to the class of non-global observables, which are known for their delicate properties and intricate calculations. Gaining deeper insights and a clearer understanding of their structure has been the subject of numerous studies over the last two decades.

Employing eikonal (soft) theory together with strong-energy ordering of emitted gluons, we have computed the fixed-order distribution of this observable up to four loops, at single-logarithmic accuracy, for both anti-$k_t$ and $k_t$ jet algorithms. Our analytical results include the full jet-radius and colour dependence. We have derived in detail the complete contributions of both NGLs and CLs and demonstrated how an optimal choice of the jet-radius, in $k_t$, may lead to the two large logarithms balancing each other out at each order and to all-orders in perturbative expansion. Additionally, we have confirmed many features identified in previous studies, such as the edge effects of both NGLs and CLs, the impact of large colour factors, and the pattern of exponentiation.

While NGLs and CLs have been addressed in prior studies, the calculations presented here, particularly for the $k_t$ jet algorithm, are distinct in that they allow these logarithms to be determined in a systematic way to any order in perturbation theory (within the above soft and energy-ordering approximations). This is enabled by the recently developed master formula for the structure of $k_t$ clustering at any order in perturbation theory \cite{Khelifa-Kerfa:2024roc}. For instance, the calculation of the four loop NGLs coefficient in k$_t$ is presented herein for the first time in the literature.

Moreover, we have derived the all-orders resummed expression for the dijet mass observable up to NLL accuracy, including both NGLs and CLs form factors. The findings align with the fixed-order results, demonstrating that the inclusion of CLs greatly reduces the effect of NGLs, thereby making the Sudakov primary form factor more reliable. This result offers an effective way of avoiding the complexities of NGLs altogether. Additionally, comparisons of the exponentiation of the analytical results with the all-orders output of numerical MC codes reveal noticeable differences between CLs and NGLs distributions. While for CLs, the analytical expression approximates the full all-orders numerical result well over a wide range of observable values, the NGLs approximation holds only for small values of the observable. This observation regarding NGLs is valid for both anti-$k_t$ and $k_t$ jet algorithms.

It is worth investigating the impact of other jet algorithms, such as Cambridge/Aachen, on the distribution of non-global observables at both fixed-order and all-orders in perturbation theory, especially since no corresponding MC codes implement the latter jet algorithm. Furthermore, it may prove useful to extend the current work to more complex QCD environments, such as Deep Inelastic Scattering (DIS) and hadron-hadron collision processes.




\bibliographystyle{spphys}
\bibliography{Refs}

\begin{thebibliography}{10}
\providecommand{\url}[1]{{#1}}
\providecommand{\urlprefix}{URL }
\expandafter\ifx\csname urlstyle\endcsname\relax
  \providecommand{\doi}[1]{DOI \discretionary{}{}{}#1}\else
  \providecommand{\doi}{DOI \discretionary{}{}{}\begingroup
  \urlstyle{rm}\Url}\fi

\bibitem{Cacciari:2008gp}
M.~Cacciari, G.P. Salam, G.~Soyez, JHEP \textbf{04}, 063 (2008).
\newblock \doi{10.1088/1126-6708/2008/04/063}

\bibitem{Catani:1993hr}
S.~Catani, Y.L. Dokshitzer, M.H. Seymour, B.R. Webber, Nucl. Phys. B
  \textbf{406}, 187 (1993).
\newblock \doi{10.1016/0550-3213(93)90166-M}

\bibitem{Ellis:1993tq}
S.D. Ellis, D.E. Soper, Phys. Rev. \textbf{D48}, 3160 (1993).
\newblock \doi{10.1103/PhysRevD.48.3160}

\bibitem{Dasgupta:2001sh}
M.~Dasgupta, G.P. Salam, Phys. Lett. B \textbf{512}, 323 (2001).
\newblock \doi{10.1016/S0370-2693(01)00725-0}

\bibitem{Dasgupta:2002bw}
M.~Dasgupta, G.P. Salam, JHEP \textbf{03}, 017 (2002).
\newblock \doi{10.1088/1126-6708/2002/03/017}

\bibitem{GehrmannDeRidder:2007hr}
A.~Gehrmann-De~Ridder, T.~Gehrmann, E.W.N. Glover, G.~Heinrich, JHEP
  \textbf{12}, 094 (2007).
\newblock \doi{10.1088/1126-6708/2007/12/094}

\bibitem{DelDuca:2016ily}
V.~Del~Duca, C.~Duhr, A.~Kardos, G.~Somogyi, Z.~Sz\H{o}r, Z.~Tr\'ocs\'anyi,
  Z.~Tulip\'ant, Phys. Rev. D \textbf{94}(7), 074019 (2016).
\newblock \doi{10.1103/PhysRevD.94.074019}

\bibitem{deFlorian:2004mp}
D.~de~Florian, M.~Grazzini, Nucl. Phys. B \textbf{704}, 387 (2005).
\newblock \doi{10.1016/j.nuclphysb.2004.10.051}

\bibitem{Becher:2008cf}
T.~Becher, M.D. Schwartz, JHEP \textbf{07}, 034 (2008).
\newblock \doi{10.1088/1126-6708/2008/07/034}

\bibitem{Abbate:2010xh}
R.~Abbate, M.~Fickinger, A.H. Hoang, V.~Mateu, I.W. Stewart, Phys. Rev. D
  \textbf{83}, 074021 (2011).
\newblock \doi{10.1103/PhysRevD.83.074021}

\bibitem{Chien:2010kc}
Y.T. Chien, M.D. Schwartz, JHEP \textbf{08}, 058 (2010).
\newblock \doi{10.1007/JHEP08(2010)058}

\bibitem{Monni:2011gb}
P.F. Monni, T.~Gehrmann, G.~Luisoni, JHEP \textbf{08}, 010 (2011).
\newblock \doi{10.1007/JHEP08(2011)010}

\bibitem{Mateu:2013gya}
V.~Mateu, G.~Rodrigo, JHEP \textbf{11}, 030 (2013).
\newblock \doi{10.1007/JHEP11(2013)030}

\bibitem{Hoang:2014wka}
A.H. Hoang, D.W. Kolodrubetz, V.~Mateu, I.W. Stewart, Phys. Rev. D
  \textbf{91}(9), 094017 (2015).
\newblock \doi{10.1103/PhysRevD.91.094017}

\bibitem{Banfi:2014sua}
A.~Banfi, H.~McAslan, P.F. Monni, G.~Zanderighi, JHEP \textbf{05}, 102 (2015).
\newblock \doi{10.1007/JHEP05(2015)102}

\bibitem{Banfi:2016zlc}
A.~Banfi, H.~McAslan, P.F. Monni, G.~Zanderighi, Phys. Rev. Lett.
  \textbf{117}(17), 172001 (2016).
\newblock \doi{10.1103/PhysRevLett.117.172001}

\bibitem{Frye:2016okc}
C.~Frye, A.J. Larkoski, M.D. Schwartz, K.~Yan,   (2016)

\bibitem{Frye:2016aiz}
C.~Frye, A.J. Larkoski, M.D. Schwartz, K.~Yan, JHEP \textbf{07}, 064 (2016).
\newblock \doi{10.1007/JHEP07(2016)064}

\bibitem{Tulipant:2017ybb}
Z.~Tulip\'ant, A.~Kardos, G.~Somogyi, Eur. Phys. J. C \textbf{77}(11), 749
  (2017).
\newblock \doi{10.1140/epjc/s10052-017-5320-9}

\bibitem{Moult:2018jzp}
I.~Moult, H.X. Zhu, JHEP \textbf{08}, 160 (2018).
\newblock \doi{10.1007/JHEP08(2018)160}

\bibitem{Bell:2018gce}
G.~Bell, A.~Hornig, C.~Lee, J.~Talbert, JHEP \textbf{01}, 147 (2019).
\newblock \doi{10.1007/JHEP01(2019)147}

\bibitem{Banfi:2018mcq}
A.~Banfi, B.K. El-Menoufi, P.F. Monni, JHEP \textbf{01}, 083 (2019).
\newblock \doi{10.1007/JHEP01(2019)083}

\bibitem{Procura:2018zpn}
M.~Procura, W.J. Waalewijn, L.~Zeune, JHEP \textbf{10}, 098 (2018).
\newblock \doi{10.1007/JHEP10(2018)098}

\bibitem{Arpino:2019ozn}
L.~Arpino, A.~Banfi, B.K. El-Menoufi, JHEP \textbf{07}, 171 (2020).
\newblock \doi{10.1007/JHEP07(2020)171}

\bibitem{Becher:2019avh}
T.~Becher, M.~Neubert, JHEP \textbf{01}, 025 (2020).
\newblock \doi{10.1007/JHEP01(2020)025}

\bibitem{Bauer:2020npd}
C.W. Bauer, A.V. Manohar, P.F. Monni, JHEP \textbf{07}, 214 (2021).
\newblock \doi{10.1007/JHEP07(2021)214}

\bibitem{Kardos:2020gty}
A.~Kardos, A.J. Larkoski, Z.~Tr\'ocs\'anyi, Phys. Lett. B \textbf{809}, 135704
  (2020).
\newblock \doi{10.1016/j.physletb.2020.135704}

\bibitem{Anderle:2020mxj}
D.~Anderle, M.~Dasgupta, B.K. El-Menoufi, J.~Helliwell, M.~Guzzi, Eur. Phys. J.
  C \textbf{80}(9), 827 (2020).
\newblock \doi{10.1140/epjc/s10052-020-8411-y}

\bibitem{Dasgupta:2022fim}
M.~Dasgupta, B.K. El-Menoufi, J.~Helliwell, JHEP \textbf{01}, 045 (2023).
\newblock \doi{10.1007/JHEP01(2023)045}

\bibitem{Duhr:2022yyp}
C.~Duhr, B.~Mistlberger, G.~Vita, Phys. Rev. Lett. \textbf{129}(16), 162001
  (2022).
\newblock \doi{10.1103/PhysRevLett.129.162001}

\bibitem{vanBeekveld:2024wws}
M.~van Beekveld, et~al.,   (2024)

\bibitem{Banfi:2002hw}
A.~Banfi, G.~Marchesini, G.~Smye, JHEP \textbf{08}, 006 (2002).
\newblock \doi{10.1088/1126-6708/2002/08/006}

\bibitem{Hatta:2013iba}
Y.~Hatta, T.~Ueda, Nucl. Phys. B \textbf{874}, 808 (2013).
\newblock \doi{10.1016/j.nuclphysb.2013.06.021}

\bibitem{Hagiwara:2015bia}
Y.~Hagiwara, Y.~Hatta, T.~Ueda, Phys. Lett. B \textbf{756}, 254 (2016).
\newblock \doi{10.1016/j.physletb.2016.03.028}

\bibitem{Hatta:2020wre}
Y.~Hatta, T.~Ueda, Nucl. Phys. B \textbf{962}, 115273 (2021).
\newblock \doi{10.1016/j.nuclphysb.2020.115273}

\bibitem{DeAngelis:2020rvq}
M.~De~Angelis, J.R. Forshaw, S.~Pl\"atzer, Phys. Rev. Lett. \textbf{126}(11),
  112001 (2021).
\newblock \doi{10.1103/PhysRevLett.126.112001}

\bibitem{Hamilton:2020rcu}
K.~Hamilton, R.~Medves, G.P. Salam, L.~Scyboz, G.~Soyez, JHEP \textbf{03}(041),
  041 (2021).
\newblock \doi{10.1007/JHEP03(2021)041}

\bibitem{vanBeekveld:2022zhl}
M.~van Beekveld, S.~Ferrario~Ravasio, G.P. Salam, A.~Soto-Ontoso, G.~Soyez,
  R.~Verheyen, JHEP \textbf{11}, 019 (2022).
\newblock \doi{10.1007/JHEP11(2022)019}

\bibitem{Banfi:2021owj}
A.~Banfi, F.A. Dreyer, P.F. Monni, JHEP \textbf{10}, 006 (2021).
\newblock \doi{10.1007/JHEP10(2021)006}

\bibitem{Banfi:2021xzn}
A.~Banfi, F.A. Dreyer, P.F. Monni, JHEP \textbf{03}, 135 (2022).
\newblock \doi{10.1007/JHEP03(2022)135}

\bibitem{Becher:2023vrh}
T.~Becher, N.~Schalch, X.~Xu, Phys. Rev. Lett. \textbf{132}(8), 081602 (2024).
\newblock \doi{10.1103/PhysRevLett.132.081602}

\bibitem{Becher:2021urs}
T.~Becher, T.~Rauh, X.~Xu, JHEP \textbf{08}, 134 (2022).
\newblock \doi{10.1007/JHEP08(2022)134}

\bibitem{FerrarioRavasio:2023kyg}
S.~Ferrario~Ravasio, K.~Hamilton, A.~Karlberg, G.P. Salam, L.~Scyboz, G.~Soyez,
  Phys. Rev. Lett. \textbf{131}(16), 161906 (2023).
\newblock \doi{10.1103/PhysRevLett.131.161906}

\bibitem{vanBeekveld:2023ivn}
M.~van Beekveld, et~al., SciPost Phys. Codeb. \textbf{2024}, 31 (2024).
\newblock \doi{10.21468/SciPostPhysCodeb.31}

\bibitem{Dasgupta:2006ru}
M.~Dasgupta, Y.~Delenda, JHEP \textbf{08}, 080 (2006).
\newblock \doi{10.1088/1126-6708/2006/08/080}

\bibitem{Banfi:2008qs}
A.~Banfi, M.~Dasgupta, Y.~Delenda, Phys. Lett. B \textbf{665}, 86 (2008).
\newblock \doi{10.1016/j.physletb.2008.05.065}

\bibitem{Banfi:2010pa}
A.~Banfi, M.~Dasgupta, K.~Khelifa-Kerfa, S.~Marzani, JHEP \textbf{08}, 064
  (2010).
\newblock \doi{10.1007/JHEP08(2010)064}

\bibitem{Khelifa-Kerfa:2011quw}
K.~Khelifa-Kerfa, JHEP \textbf{02}, 072 (2012).
\newblock \doi{10.1007/JHEP02(2012)072}

\bibitem{Kerfa:2012yae}
K.K. Kerfa, {QCD resummation for high-$p_T$ jet shapes at hadron colliders}.
\newblock Ph.D. thesis, Manchester U. (2012)

\bibitem{Dasgupta:2012hg}
M.~Dasgupta, K.~Khelifa-Kerfa, S.~Marzani, M.~Spannowsky, JHEP \textbf{10}, 126
  (2012).
\newblock \doi{10.1007/JHEP10(2012)126}

\bibitem{Bouaziz:2022tik}
H.~Bouaziz, Y.~Delenda, K.~Khelifa-Kerfa, JHEP \textbf{10}, 006 (2022).
\newblock \doi{10.1007/JHEP10(2022)006}

\bibitem{Schwartz:2014wha}
M.D. Schwartz, H.X. Zhu, Phys. Rev. D \textbf{90}(6), 065004 (2014).
\newblock \doi{10.1103/PhysRevD.90.065004}

\bibitem{Caron-Huot:private}
S.~Caron-Huot,   (2015).
\newblock {private communication}

\bibitem{Benslama:2020wib}
H.~Benslama, Y.~Delenda, K.~Khelifa-Kerfa, A.M. Ibrahim, Phys. Part. Nucl.
  Lett. \textbf{18}(1), 5 (2021).
\newblock \doi{10.1134/S1547477121010039}

\bibitem{Khelifa-Kerfa:2015mma}
K.~Khelifa-Kerfa, Y.~Delenda, JHEP \textbf{03}, 094 (2015).
\newblock \doi{10.1007/JHEP03(2015)094}

\bibitem{Benslama:2023gys}
H.~Benslama, Y.~Delenda, K.~Khelifa-Kerfa, Phys. Lett. B \textbf{840}, 137903
  (2023).
\newblock \doi{10.1016/j.physletb.2023.137903}

\bibitem{Khelifa-Kerfa:2024udm}
K.~Khelifa-Kerfa, JHEP \textbf{10}, 079 (2024).
\newblock \doi{10.1007/JHEP10(2024)079}

\bibitem{Appleby:2002ke}
R.B. Appleby, M.H. Seymour, JHEP \textbf{12}, 063 (2002).
\newblock \doi{10.1088/1126-6708/2002/12/063}

\bibitem{Appleby:2003sj}
R.B. Appleby, M.H. Seymour, JHEP \textbf{09}, 056 (2003).
\newblock \doi{10.1088/1126-6708/2003/09/056}

\bibitem{Banfi:2005gj}
A.~Banfi, M.~Dasgupta, Phys. Lett. \textbf{B628}, 49 (2005).
\newblock \doi{10.1016/j.physletb.2005.08.125}

\bibitem{Delenda:2006nf}
Y.~Delenda, R.~Appleby, M.~Dasgupta, A.~Banfi, JHEP \textbf{0612}, 044 (2006).
\newblock \doi{10.1088/1126-6708/2006/12/044}

\bibitem{Kelley:2012kj}
R.~Kelley, J.R. Walsh, S.~Zuberi, JHEP \textbf{09}, 117 (2012).
\newblock \doi{10.1007/JHEP09(2012)117}

\bibitem{Kelley:2012zs}
R.~Kelley, J.R. Walsh, S.~Zuberi.
\newblock Disentangling clustering effects in jet algorithms (2012).
\newblock \urlprefix\url{https://arxiv.org/abs/1203.2923}

\bibitem{Ziani:2021dxr}
N.~Ziani, K.~Khelifa-Kerfa, Y.~Delenda, Eur. Phys. J. C \textbf{81}, 570
  (2021).
\newblock \doi{10.1140/epjc/s10052-021-09379-z}

\bibitem{Delenda:2012mm}
Y.~Delenda, K.~Khelifa-Kerfa, JHEP \textbf{09}, 109 (2012).
\newblock \doi{10.1007/JHEP09(2012)109}

\bibitem{VHjet_kt_4loop}
K.~Khelifa-Kerfa, {``NGLs up to four-loops for V/H+jet processes in k$_t$
  algorithm"}.
\newblock {in progress}

\bibitem{Dokshitzer:1997in}
Y.L. Dokshitzer, G.D. Leder, S.~Moretti, B.R. Webber, JHEP \textbf{08}, 001
  (1997).
\newblock \doi{10.1088/1126-6708/1997/08/001}

\bibitem{Wobisch:1998wt}
M.~Wobisch, T.~Wengler, in \emph{{Workshop on Monte Carlo Generators for HERA
  Physics (Plenary Starting Meeting)}} (1998), pp. 270--279

\bibitem{Becher:2023znt}
T.~Becher, J.~Haag, JHEP \textbf{01}, 155 (2024).
\newblock \doi{10.1007/JHEP01(2024)155}

\bibitem{Delenda:2015tbo}
Y.~Delenda, K.~Khelifa-Kerfa, Phys. Rev. D \textbf{93}(5), 054027 (2016).
\newblock \doi{10.1103/PhysRevD.93.054027}

\bibitem{Kelley:2011tj}
R.~Kelley, M.D. Schwartz, H.X. Zhu.
\newblock Resummation of jet mass with and without a jet veto (2011).
\newblock \urlprefix\url{https://arxiv.org/abs/1102.0561}

\bibitem{Khelifa-Kerfa:2024roc}
K.~Khelifa-Kerfa.
\newblock The analytical structure of k$_t$ clustering to any order (2024).
\newblock \urlprefix\url{https://arxiv.org/abs/2409.14029}

\bibitem{Catani:1996jh}
S.~Catani, M.H. Seymour, Phys. Lett. \textbf{B378}, 287 (1996).
\newblock \doi{10.1016/0370-2693(96)00425-X}

\bibitem{Cacciari:2011ma}
M.~Cacciari, G.P. Salam, G.~Soyez, Eur. Phys. J. C \textbf{72}, 1896 (2012).
\newblock \doi{10.1140/epjc/s10052-012-1896-2}

\bibitem{Cacciari:2005hq}
M.~Cacciari, G.P. Salam, Phys. Lett. \textbf{B641}, 57 (2006).
\newblock \doi{10.1016/j.physletb.2006.08.037}

\bibitem{Catani:1992ua}
S.~Catani, L.~Trentadue, G.~Turnock, B.R. Webber, Nucl. Phys. \textbf{B407}, 3
  (1993).
\newblock \doi{10.1016/0550-3213(93)90271-P}

\end{thebibliography}

\end{document}